\documentclass[a4paper,fleqn,usenatbib]{mnras}
\usepackage[T1]{fontenc}
\usepackage{ae,aecompl}
\usepackage{amsmath}
\usepackage{amsfonts}
\usepackage{amssymb}
\usepackage{graphicx}
\def\mnras{MNRAS}
\def\aap{A$\&$A}
\def\nat{Nature}
\def\apj{ApJ}
\def\aj{AJ}
\def\icarus{Icarus}

\newcommand{\mmsn}{{\rm MMSN}}
\newcommand{\me}{\, {\rm M}_{\oplus}}
\newcommand{\au}{\, {\rm au}}
\title[Formation of compact planetary systems]{On the formation of compact planetary systems \emph{via} concurrent core accretion and migration}
\author[G. A. L. Coleman and R. P. Nelson]{Gavin A. L. Coleman\thanks{Email: g.coleman@qmul.ac.uk} \& Richard P. Nelson \\
Astronomy Unit, Queen Mary University of London, Mile End Road, London, E1 4NS, U.K.}
\date{Accepted 2016 January 14. Received 2016 January 13; in original form 2015 November 26}
\begin{document}
\label{firstpage}
\pagerange{\pageref{firstpage}--\pageref{lastpage}}
\maketitle
\begin{abstract}
We present the results of planet formation N-body simulations based on a comprehensive physical model that includes planetary mass growth through mutual embryo collisions and planetesimal/boulder accretion, viscous disc evolution, planetary migration and gas accretion onto planetary cores. The main aim of this study is to determine which set of model parameters leads to the formation of planetary systems that are similar to the compact low mass multi-planet systems that have been discovered by radial velocity surveys and the \emph{Kepler} mission. We vary the initial disc mass, solids-to-gas ratio and the sizes of the boulders/planetesimals, and for a restricted volume of the parameter space we find that compact systems containing terrestrial planets, super-Earths and Neptune-like bodies arise as natural outcomes of the simulations. Disc models with low values of the solids-to-gas ratio can only form short-period super-Earths and Neptunes when small planetesimals/boulders provide the main source of accretion, since the mobility of these bodies is required to overcome the local isolation masses for growing embryos. The existence of short-period super-Earths around low metallicity stars provides strong evidence that small, mobile bodies (planetesimals, boulders or pebbles) played a central role in the formation of the observed planets.
\end{abstract}
\section{Introduction}
Both radial velocity \citep{2011arXiv1109.2497M} and transit surveys have shown conclusively that systems of low mass planets are common around main sequence stars, with the \emph{Kepler} mission in particular providing some striking examples of short period compact multi-planet systems 
\citep{Lissauer2011, Fabrycky2014}. The most recent release of \emph{Kepler} data contains over 4700 planet candidates, and more than 700 multi-planet systems \citep{Mullally2015}. Approximately 3000 systems show just a single transiting planet candidate, with orbital periods in the range $0.5 \le P \le 500$ days.

Analysis of the systems properties provides useful insight for understanding how these planets formed and evolved. One noticeable feature of the multi-systems is the paucity of first order mean motion resonances. The period ratio distribution shows features in the vicinity of the 2:1 and 3:2 resonances, suggesting that they have been dynamically important in the past, but relatively few systems are actually in a strict mean motion resonance \citep{Fabrycky2014}. Examples of systems of small planets that are in or very close to resonance, including 3 body resonances or resonant chains, include Kepler 50 (6:5), Kepler 60 (5:4, 4:3) \citep{Steffen2012}, Kepler 221 (displays a 3 body resonance) \citep{Fabrycky2014}. In general the compact multi-planet systems appear to be composed of terrestrial planets, super-Earths and Neptune-like bodies. Mass estimates based on both radial velocity and transit timing variations suggest that there is a strong diversity in the mean densities of these objects, with some being rocky, some appearing to have a mixture of rock and water, and others being of very low density indicating the presence of significant fractions of H/He \citep{Lissauer2011, WuLithwick2013, Marcy2014, Jontof-Hutter2015}. Kepler 36 provides an example where a pair of neighbouring planets orbiting close to the 7:6 resonance have dramatically different densities, characteristic of a rocky terrestrial inner body and an outer mini-Neptune \citep{Carter2012}. One of the most interesting facts to emerge from the data is the presence of low mass planetary systems around stars with a broad range of metallicities, including stars whose iron contents are factors of $\sim 3$ smaller than the solar abundance \citep{Buchhave2014}, a result that is supported by radial velocity discoveries of planets around metal-poor M dwarfs, such as Kapteyn's star \citep{Anglada-Escude2014}.
 
A number of ideas have been put forward to explain the formation and early evolution of the compact \emph{Kepler} and radial velocity systems, which in cases such as Gliese 581 and HD 69830 appear to contain in excess of $\sim 30 \me$ of solid material within a few tenths of an au \citep{Lovis2006,Udry2007}. This concentration of solids close to the star led to classical core accretion models combined with disc driven migration being developed using population synthesis codes \citep{Alibert2006}. More recent population synthesis calculations that also include prescriptions for planet-planet interactions have also been presented \citep{IdaLin2010}. N-body simulations, combined with either hydrodynamic simulations or analytic prescriptions for migration and eccentricity/inclination damping of planetary growth, have also been used to examine the origins of such
systems \citep{CresswellNelson2006, cressnels, TerquemPapaloizou2007, McNeilNelson2009, McNeilNelson2010, Hellary, Cossou, ColemanNelson14, Hands14}. A common outcome of these N-body simulations is the formation of resonant convoys of planets in the presence of convergent migration, an outcome that is not reflected in the \emph{Kepler} systems. Various ideas have been put forward to explain why the resonances may be unstable, including tidal eccentricity damping followed by separation of the resonance for short period systems \citep{TerquemPapaloizou2007}, stochastic migration due to local turbulence \citep{Adams2008,ReinPapaloizou2009,Rein2012} - a process that is likely to only operate close to the star where the disc can be thermally ionised \citep{UmebayashiNakano1988, DeschTurner2015}, resonance breaking due to overstable librations  \citep{GoldreichSchlichting2014}, orbital repulsion due to nonlinear spiral wave damping in planet coorbital regions \citep{BaruteauPapaloizou2013, Podlewska-Gaca2012}.  

The paucity of mean motion resonances in the \emph{Kepler} data has led to suggestions that the compact systems formed \emph{in situ} through giant impacts, akin to the final stages of accumulating the terrestrial planets \citet{ChambersWetherill1998}, after the concentration of small planetesimals in the inner disc followed by their growth into planetary embryos \citep{HansenMurray2012}. Although this model has some success in generating non resonant multiple planet systems with inclinations that are in good agreement with \emph{Kepler} systems, there are difficulties in explaining how such large amounts of solids become concentrated in the inner disc, and the model fails to reproduce the numbers of single transiting planets detected by \emph{Kepler} \citep{HansenMurray2012}. An alternative \emph{in situ} model has been proposed by \citet{ChatterjeeTan2014} where pebbles/boulders concentrate and form a planet at the pressure maximum generated at the interface between the inner turbulent region of the disc and the dead zone, and exterior planets are spawned in succession by the disc being eroded outwards when the planets reach gap forming masses. While this model may be able to explain some systems, it is not clear that such a model can work for systems such as Kepler 444 and Kepler 186 where the planet masses are likely to be too small to form gaps, or for planetary systems in which the innermost planets orbit further from their stars than the fully active regions are expected to extend.

In this paper we present the results from a suite of N-body simulations using an updated version of the planet formation and protoplanetary disc model presented in \citet{ColemanNelson14}, hereafter referred to as CN14. A basic assumption of the model is that the protoplanetary disc contains a population of planetary embryos distributed across a wide range of orbital radii (between 1 - 20 au), which grow through the accretion of boulders or planetesimals, and through mutual collisions, and can accrete gas from the nebula when they reach masses $\ge 3 \me$. We refer to this as a \emph{distributed core accretion model}, in contrast to one where smaller numbers of embryos might form at specific disc locations such as pressure maxima through the trapping and accumulation of solids. We include the most up-to-date prescriptions for migration, self-consistent evolution of the viscous disc, and disc removal by a photoevaporative wind on multi-Myr time scales. The main updates on CN14 include placing the inner boundary of the computational domain close to the star so that we can simulate planets that migrate to regions with orbital periods down to 1 day, addition of an active turbulent region (mimicked as a simple increase in viscosity) where disc temperatures exceed 1000~K, and a magnetospheric cavity close to the star into which planets can migrate. The aim of this work is simply to examine whether or not such a comprehensive model of planet formation is able to produce planetary systems that are similar to those that have been observed, and if so under which set of conditions (disc mass, metallicity, planetesimal/boulder sizes) do these systems form. We emphasise that this is not population synthesis. No attempt is made to select initial conditions from a distribution of possibilities constrained by observations. We are not aiming to reproduce the frequency with which certain types of planetary systems arise, but instead to just examine which conditions allow \emph{Kepler}-like compact systems to form subject to our model assumptions. We find that the simulations produce a broad range of outcomes that correlate strongly with the amount of solids present initially in the disc, and with the sizes of the boulders/planetesimals that provide the primary feedstock for planetary growth. Short-period compact multi-systems containing both resonant and non resonant planet pairs are one particular outcome of the runs, and these arise within a restricted range of the parameter space that we consider.

The paper is organised as follows.
We present the physical model and numerical methods in Section \ref{sec:meth}, and our simulation results in Section \ref{sec:results}.
We compare our results with observations in Section \ref{sec:comparison}, and we draw conclusions in Section \ref{sec:discussion}.

\section{Physical model and numerical methods}
\label{sec:meth}
The N-body simulations presented here were performed using the Mercury-6 symplectic integrator \citep{Chambers}, 
adapted to include the disc models and physical processes described below.
We use an updated version of the physical model described in CN14. The main elements of this model are described below, and the implemented updates are outlined in the following subsections. The basic model consists of 52 
protoplanets, orbiting within a swarm of thousands of boulders or planetesimals, 
embedded in a gaseous protoplanetary disc, all orbiting around a solar mass star.
For each simulation we adopt a single size for the boulders or planetesimals.
We define objects of radius $R_{\rm pl}=10$~m to be \emph{boulders} and
objects of radius $R_{\rm pl} \ge 100$~m to be \emph{planetesimals}. These various sized objects
differ from each other and from \emph{protoplanets} or \emph{planetary embryos}
because they experience gas drag forces that vary with the size.

\subsection{Recap on the CN14 model}
The basic model from CN14 is comprised of the following elements:

\noindent
(i) The standard diffusion equation for a 1D viscous $\alpha$-disc model is solved \citep{Shak,Lynden-BellPringle1974}. 
Temperatures are calculated by balancing blackbody cooling against viscous heating and 
stellar irradiation. In the presence of a giant planet tidal torques can open a gap in the disc.
 
\noindent
(ii) The final stages of disc removal occur through a photoevaporative wind.
A standard photoevaporation model is used for most of the disc evolution \citep{Dullemond},
but if a large cavity forms in the presence of a gap forming planet, direct photoevaporation of 
the disc is switched on if the planet sits outside of the innermost radius from where the thermally-driven wind can be 
launched \citep{Alexander09}, as outlined in Section 5 of CN14.
 
\noindent
(iii) Planetesimals and boulders orbiting in the disc experience aerodynamic drag.
 
\noindent
(iv) We use the torque formulae from \citet{pdk10,pdk11} to simulate type I migration
due to Lindblad and corotation torques. Corotation torques arise from both entropy
and vortensity gradients in the disc, and the possible saturation of these torques is 
included in the simulations. The influence of eccentricity and inclination on the
migration torques, and on eccentricity and inclination damping are included
\citep{Fendyke,cressnels}.

\noindent
(v) Type II migration of planets is included via the impulse approximation of 
\citet{LinPapaloizou86} if they reach the gap opening mass.

\noindent
(vi) Gas envelope accretion from the surrounding disc occurs for planets
whose masses exceed $3 \me$ using fits to detailed 1D models from \citet{Movs}. 
Gas accretion occurs at the local viscous supply rate for gap forming planets. Type II migration, and gas accretion rates through the gap, have been calibrated against hydrodynamic simulations as described in CN14.

\noindent
(vii) The effective capture radius of protoplanets accreting planetesimals is enhanced
by atmospheric drag as described in \citet{Inaba}.

\subsection{Model improvements and additions}
\subsubsection{Active turbulent region}
Fully developed magnetohydrodynamic turbulence is expected to arise in regions of the disc
where the temperature exceeds 1000~K \citep{UmebayashiNakano1988, DeschTurner2015}. 
To account for the increased turbulent stress we increase the viscous $\alpha$ parameter 
when the temperature rises above 1000~K using the prescription
\begin{equation}
\alpha(r) = \left\{ \begin{array}{ll}
2\times10^{-3} & r > r_s, \\
2\times10^{-3} + 4\times10^{-3} \\
\times\left(\tanh\left(\dfrac{3(r_s-r-5H(r))}{5H(r)}\right)+1\right) & r \le r_s, \\
\end{array} \right.
\end{equation}
where $r_s$ represents the outermost radius with temperature greater than 1000 K, and $H(r)$ 
is the local disc scale height. This transition leads to a maximum $\alpha=10^{-2}$ in
the hottest parts of the disc sitting within $\sim 0.5 \au$ from the star at the beginning
of the simulations.

\subsubsection{Magnetospheric cavity and inner boundary}
\label{sec:magnetospheric_cavity}
A rotating star with a strong dipole magnetic field may create an inner disc cavity 
through magnetic torques repelling the disc, and this can provide an effective
mechanism for preventing planets migrating into their host stars \citep[e.g.][]{Lin_Boden_96}.
We include a cavity in our simulations by assuimg that the outer edge of the cavity
is truncated at $0.05 \au$, corresponding to an orbital period of $\sim 4$ days,
in agreement with the spin periods of numerous T Tauri stars \citep{Herbst_Mundt_05}.
Planets are able to migrate into this region through either type I or type II migration. A planet that has not reached the local gap opening mass halts its migration once it reaches the cavity edge (the assumption here is that strong corotation torques will stop its migration,\textbf{ as shown for migrating circumbinary planets \citep{PierensNelson07}, and those migrating in towards a single central star \citep{Benitez-Llambay11}}). A gap forming planet continues to migrate into the cavity until it reaches the 2:1 orbital resonance with the cavity outer edge, at which point disc torques are switched off. This resonance is located at $\sim 0.0315 \au$ from the star. 
It should be noted that a second planet entering the cavity can nudge a planet sitting at the 2:1 resonance location onto a shorter period orbit. The inner boundary of the computational domain is located
just inside $0.02 \au$ (corresponding to a $\sim 1$ day orbit period). Any planets whose semimajor axes are smaller than the boundary
radius are removed from the simulation and are assumed to have hit the star. We note 
that the inner boundary adopted in CN14 corresponded to an orbital period of 20 days.

A summary of the disc and stellar parameters adopted in all simulations is given in Table~\ref{tab:modelparam}.

\begin{table}
\begin{tabular}{lc}
\hline
Parameter & Value\\
\hline
Disc inner boundary & 0.02 au\\
Cavity outer boundary & 0.05 au\\
Disc outer boundary & 40 au\\
Number of cells & 1000\\
$\Sigma_{\rm g}$(1 au) & $1731$ g\,${\rm cm}^{-2}$\\
Stellar Mass & $1\rm M_{\bigodot}$\\
$R_{\rm S}$ & $2 \rm R_{\bigodot}$\\
$T_{\rm S}$ & 4280 K\\
\hline
\end{tabular}
\caption{Disc and stellar model parameters}
\label{tab:modelparam}
\end{table}

\subsubsection{Opacity}
We make a small change to the opacity prescription used in CN14 by assuming that half of the disc solids are
in submicron sized dust particles, with the remainder being in planetary embryos and planetesimals/boulders.
The opacity used to calculate the thermal diffusion timescale in the disc is thus multiplied by the
factor $F_{\rm opacity}= 1/2 \times {\rm M}_{\rm ratio}$ where ${\rm M_{\rm ratio}}$ is the ratio of the disc metallicity
to the solar metallicity. $F_{\rm opacity} =1/2$ for a disc with solar metallicity, $1/4$ for a
disc with half the solar metallicity, and 1 for a disc with twice solar metallicity. This modification of the opacity affects both the equilibrium disc temperature and estimates for when the corotation torques acting on planets saturate.

\subsubsection{Gas envelope accretion}
\begin{figure}
\includegraphics[scale=0.45]{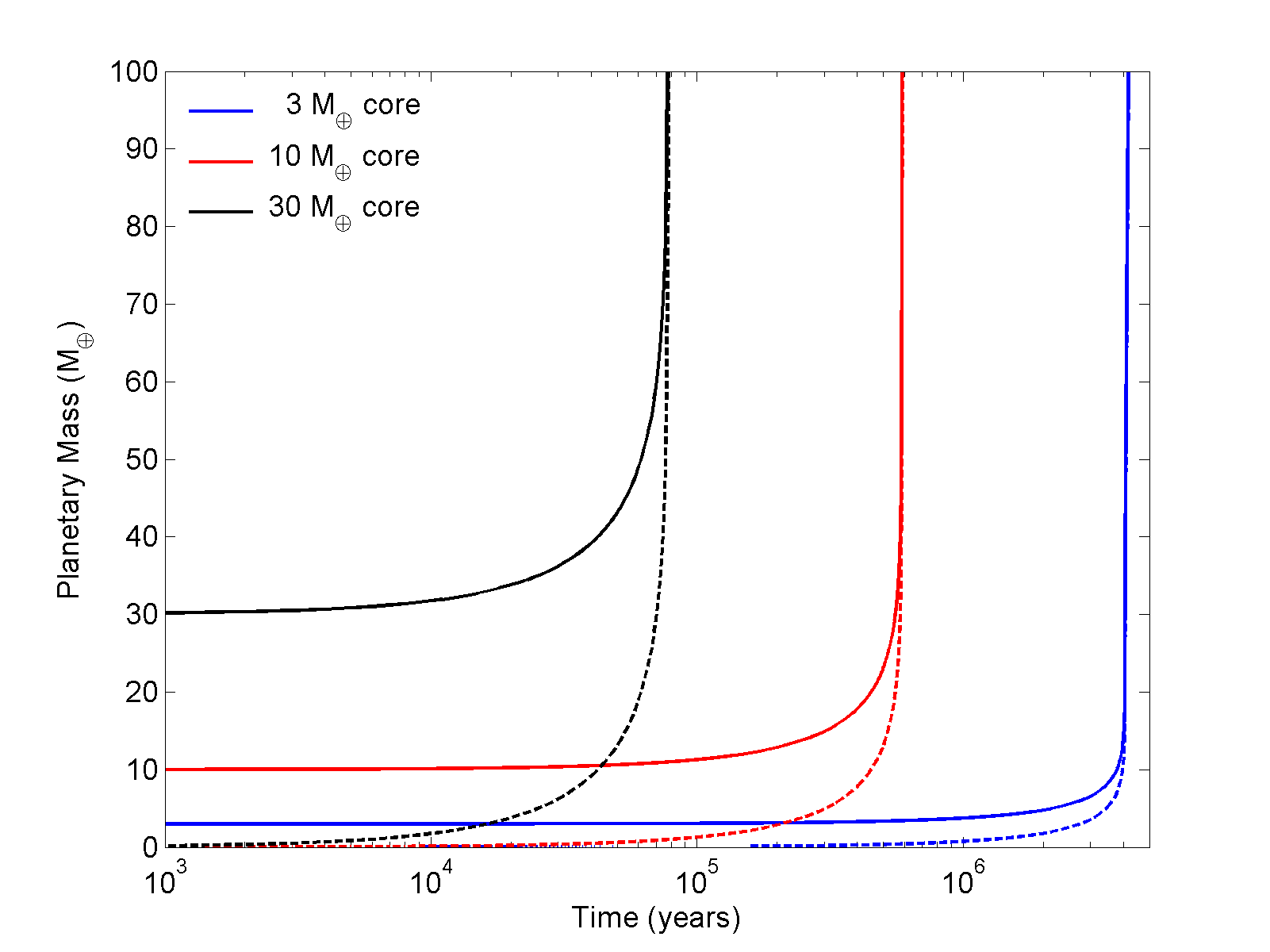}
\caption{Gas accretion onto 3, 10, and 30 $\me$ cores versus time at 5 $\au$. Solid lines denote total mass, whilst dotted lines denote the envelope mass.}
\label{fig:newgrowth}
\end{figure}

Once a protoplanet grows to a mass that exceeds 3 M$_{\oplus}$ it starts to accrete
a gaseous envelope.  We have improved on the fits to the 1D giant planet formation
models of \citet{Movs} used in CN14. In units of Earth masses and Myr, our improved 
scheme gives a gas accretion rate of:

\begin{equation}
\label{eq:mov}
\dfrac{dm_{\rm ge}}{dt} = \dfrac{4.5}{96.65}\exp{\left(\dfrac{m_{\rm ge}}{22}\right)}m_{\rm core}^{2.4}\exp{\left(\dfrac{m_{\rm ge}}{m_{\rm core}}\right)}.
\end{equation}
This scheme allows for the continuation of core growth after a gaseous envelope has been acquired, 
while allowing the rate of envelope accretion to adapt to the varying core and envelope mass.
Figure \ref{fig:newgrowth} shows gas accretion onto 3, 10 and 30 $\me$ cores without the influence of 
migration or core growth.
These are similar to the models in \citet{Hellary} and CN14, but are in better agreement with 
the models presented by \citet{Movs}.
Ideally, we would incorporate self-consistent models of gas envelope accretion in
the simulations, but unfortunately this is too expensive computationally to run within our current model.
While our fits to the \citet{Movs} models allows gas accretion to occur at the rates prescribed in that work, 
these fits do not change according to the local conditions in the disc, or to a time varying planetesimal 
accretion rate. This is something that we will address in future work.

The gas accretion rate given by eqn.~\ref{eq:mov} is valid until the planet forms a gap within the disc, after which the gas accretion rate switches to either the value obtained  from eqn.~\ref{eq:mov} or the viscous supply rate given by
\begin{equation}
\dfrac{dm_{ge}}{dt}=3\pi\nu\Sigma_g,
\end{equation}
whichever is smaller. Here $\Sigma_g$ and $\nu$ are the  surface density and viscosity of the gas that sits at a distance of 10 Hill radii exterior to the planet's location.
This prescription is chosen because the planet sits in a deep gap and so the supply rate of gas must be evaluated at a location in the disc that sits outside the fully evacuated gap region.
The precise value that is quoted here was determined in Section 5.2 of CN14 where different evaluation distances were tested against 2D hydrodynamic simulations, and 10 Hill radii showed the best agreement.
We note that our gas accretion routine conserves mass. Gas that is accreted onto the planet is removed from the surrounding disc.

\subsubsection{Aerodynamic drag}
Solid bodies experience aerodynamic drag, reducing semimajor axes whilst simultaneously damping eccentricities and inclinations.
Stoke's drag is applied to planetesimals/boulders \citep{Adachi} when the size of the body is greater than twice the molecular mean free path ($\lambda$). This switches to Epstein drag  when the mean free path exceeds roughly half the planetesimal size \citep{Weidenschilling_77}.
Here $\lambda$ is given by:
\begin{equation}
\lambda = \dfrac{\mu m_{\rm H}}{\sigma\rho_{\rm g}}
\end{equation}
where $\sigma$ is the collision cross section, $\mu$ is the gas mean molecular weight, and $\rho_{\rm g}$ is the local gas density.
When the planetesimal size is greater than $\frac{9}{4}\lambda$ we use Stokes' drag law given as:
\begin{equation}
\textbf{F}_{st} = m_{\rm pl} \left(\dfrac{-3 \rho_{\rm g} C_{\rm D}}{8\rho_{\rm pl}R_{\rm pl}}\right) 
v_{\rm rel}\textbf{v}_{\rm rel}
\end{equation}
Here, a subscript `${\rm pl}$' corresponds to planetesimals, $\rho_{\rm pl}$ is the internal 
density of planetesimals, $R_{\rm pl}$ is the planetesimal radius, and $v_{\rm rel}$ is the 
relative velocity between the gas and planetesimals.
$C_D$ is the dimensionless drag coefficient, taken as a function of the Reynolds number ($R_e$) given below
\begin{equation}
C_D = \left\{ \begin{array}{lll}
24 R_e^{-1} & R_e < 1 \\
24 R_e^{-0.6} & 1 \le R_e < 800 \\
0.44 & R_e > 800
\end{array} \right.
\end{equation}
When the planetesimal size is equal to $9\lambda/4$, both drag regimes are equal, thus we transition to the Epstein drag law given as:
\begin{equation}
\textbf{F}_{ep} = m_{\rm pl} \left(\dfrac{\rho}{\rho_{\rm pl}R_{\rm pl}}\right)v_{\rm rel}c_s
\end{equation}
When the planetesimal size is smaller than $9\lambda/4$ we only use the Epstein drag law.

\begin{table*}
\begin{tabular}{lcccc}
\hline
Simulation & Disc mass & Metallicity   & Planetesimal radius & Formation behaviour\\
           & (MMSN)    & (solar value) & (km) & (A+B) \\
\hline
K10.50.01A, K10.50.01B & 1 & 0.5 & 0.01 & LPG\\
K10.50.1A, K10.50.1B & 1 & 0.5 & 0.1 & LPG\\
K10.51A, K10.51B & 1 & 0.5 & 1 & LPG\\
K10.510A, K10.510B & 1 & 0.5 & 10 & LPG\\
K110.01A, K110.01B & 1 & 1 & 0.01 & MGM\\
K110.1A, K110.1B & 1 & 1 & 0.1 & LPG\\
K111A, K111B & 1 & 1 & 1 & LPG\\
K1110A, K1110B & 1 & 1 & 10 & LPG\\
K120.01A, K120.01B & 1 & 2 & 0.01 & GFSM\\
K120.1A, K120.1B & 1 & 2 & 0.1 & MGM\\
K121A, K121B & 1 & 2 & 1 & LPG\\
K1210A, K1210B & 1 & 2 & 10 & LPG\\

K1.50.50.01A, K1.50.50.01B & 1.5 & 0.5 & 0.01 & MGM\\
K1.50.50.1A, K1.50.50.1B & 1.5 & 0.5 & 0.1 & LPG\\
K1.50.51A, K1.50.51B & 1.5 & 0.5 & 1 & LPG\\
K1.50.510A, K1.50.510B & 1.5 & 0.5 & 10 & LPG\\
K1.510.01A, K1.510.01B & 1.5 & 1 & 0.01 & GFSM\\
K1.510.1A, K1.510.1B & 1.5 & 1 & 0.1 & MGM\\
K1.511A, K1.511B & 1.5 & 1 & 1 & LPG\\
K1.5110A, K1.5110B & 1.5 & 1 & 10 & LPG\\
K1.520.01A, K1.520.01B & 1.5 & 2 & 0.01 & GFSM\\
K1.520.1A, K1.520.1B & 1.5 & 2 & 0.1 & GFSM\\
K1.521A, K1.521B & 1.5 & 2 & 1 & LPG\\
K1.5210A, K1.5210B & 1.5 & 2 & 10 & LPG\\

K20.50.01A, K20.50.01B & 2 & 0.5 & 0.01 & MGM\\
K20.50.1A, K20.50.1B & 2 & 0.5 & 0.1 & MGM\\
K20.51A, K20.51B & 2 & 0.5 & 1 & LPG\\
K20.510A, K20.510B & 2 & 0.5 & 10 & LPG\\
K210.01A, K210.01B & 2 & 1 & 0.01 & GFSM\\
K210.1A, K210.1B & 2 & 1 & 0.1 & MGM\\
K211A, K211B & 2 & 1 & 1 & LPG\\
K2110A, K2110B & 2 & 1 & 10 & LPG\\
K220.01A, K220.01B & 2 & 2 & 0.01 & GFSM\\
K220.1A, K220.1B & 2 & 2 & 0.1 & GFSM\\
K221A, K221B & 2 & 2 & 1 & MGM\\
K2210A, K2210B & 2 & 2 & 10 & LPG\\

\hline
\end{tabular}
\caption{Simulation parameters with formation behaviours as follows: LPG - Limited Planetary Growth, MGM - Moderate Growth and Migration, GFSM - Migrating Giants.}
\label{tab:simparam}
\end{table*}

\begin{table*}
\centering
\begin{tabular}{lcccc}
\hline
Classification & Mass & Rock $\%$ & Ice $\%$ & Gas $\%$ \\
\hline
Rocky terrestrial & $m_p<3M_{\oplus}$ & $>70\%$ & $<30\%$ & $0\%$\\
Water-rich terrestrial & $m_p<3M_{\oplus}$ & $<70\%$ & $>30\%$ & $0\%$\\
Rocky super-Earth & $3M_{\oplus}\leq m_p<10M_{\oplus}$ & $>60\%$ & $<30\%$ & $<10\%$\\
Water-rich super-Earth & $3M_{\oplus}\leq m_p<10M_{\oplus}$ & N/A & $>30\%$ & $<10\%$\\
Mini-Neptune & $3M_{\oplus}\leq m_p<10M_{\oplus}$ & N/A & N/A & $>10\%$\\
Gas-rich Neptune & $10M_{\oplus}\leq m_p<35M_{\oplus}$ & N/A & N/A & $>10\%$\\
Gas-poor Neptune & $10M_{\oplus}\leq m_p<35M_{\oplus}$ & N/A & N/A & $<10\%$\\
Gas-dominated giant & $m_p\geq35M_{\oplus}$ & N/A & N/A & $>50\%$\\
Core-dominated giant & $m_p\geq35M_{\oplus}$ & N/A & N/A & $<50\%$\\
\hline
\end{tabular}
\caption{Planetary classification parameters based on their composition and the
mass fraction of their gaseous envelope. Note that water-rich planets are so-called
because they accrete water ice in solid form that originates from beyond the snow-line.}
\label{tab:plcompo}
\end{table*}

\subsection{Initial conditions}
All simulations were run for 10~Myr, allowing the systems of formed planets to continue evolving through scattering and collisions after the dispersal of the protoplanetary discs. 
A run time of 10~Myr is insufficient for accretion between embryos orbiting at large distances to reach completion,
and some of our simulations were halted when systems of planets on longer period orbits were still evolving. This is
unavoidable for systems in which large scale migration leads to the formation of short period planets, with longer period planets remaining at larger semimajor axes, 
since the time steps become prohibitively short for Gyr-run times to be achieved. For this reason, most of
our discussion will focus on the short period systems that arise in the simulations as these are dynamically much
more mature than the longer period planets.

The runs were all initiated with 52 planetary embryos, of mass $0.1\me$, separated by 10 mutual Hill radii, and with 
semimajor axes between 1 and 20 $\au$. These were embedded in a swarm of thousands of planetesimals/boulders, that were 
distributed with semimajor axes between 0.5 and 20 $\au$, and with masses either 10, 20 or 50 times smaller than the embryos, depending on the metallicity of the system. (This varying mass ratio between embryos and planetesimals was implemented to keep the numbers of planetesimals at a number that allowed the simulations to run on reasonable times scales. Between 3000 and 8000 planetesimals/boulders were used and run times for the individual simulations varied between 3 and 9 months.)
The \textbf{effective} physical radii of planetesimals were set to either 10~m, 100~m, 1~km and 10~km, such that the primary feedstock of the accreting protoplanets ranged from being boulders to being large planetesimals whose evolution differed principally because of the strengths of the gas drag forces that they experienced.
\textbf{Planetesimals/boulders in our simulations represent a larger group of particles, with realistic masses depending on their physical radii, whose averaged orbits allow them to be approximated as a single massive super-particle with an effective physical radius.}
Eccentricities and inclinations for protoplanets and planetesimals/boulders were randomized according to a Rayleigh distribution, with scale parameters $e_0=0.01$ and $i_0=0.25^{\circ}$, respectively.

Collisions between protoplanets and other protoplanets or planetesimals resulted in perfect sticking.
We neglect planetesimal-planetesimal interactions and collisions in our simulations for reasons of computational speed.

The gas disc masses simulated were 1, 1.5 and 2 times the mass of the minimum mass solar nebula (MMSN) \citep{Hayashi}. We also vary the disc metallicity so that the initial solids-to-gas mass ratios are equal to 0.5, 1 and 2 times the solar value for the different models.
We smoothly increase the mass of solids exterior to the snow line by a factor of 4, as described in \citet{Hellary}.
We track the changes in planetary compositions throughout the simulations, as planets can accrete material originating either interior or exterior to the snow line.

Combining the three different gas disc masses, the three values of metallicity/solids-to-gas mass ratio, and the four different planetesimal/boulder sizes gives a total of 36 parameter variations.
We ran two instances of each parameter set, where only the random number seed to generate initial particle positions 
and velocities was changed, giving a total of 72 simulations. The full set of run parameters are detailed in Table \ref{tab:simparam}.

\section{Results}
\label{sec:results}
In order to provide context for our N-body simulations, we begin discussion of the 
results by describing the general evolution of the disc models, and the orbital evolution 
of the protoplanets and planetesimals. We then recap the main results obtained in CN14
before describing the results of the new simulations. We divide the results of the new
runs into three distinct categories: \emph{limited planetary growth (LPG)}; 
\emph{moderate growth and migration (MGM)}; \emph{giant formation and significant migration (GFSM)}.
For each category, we present the details of one or two representative runs, with
Table \ref{tab:simparam} listing the category for each run. Runs that displayed 
\emph{limited planetary growth} resulted in no planet masses growing above $3 \me$ during the gas disc life time (and hence the amount of type I migration was also modest), although further growth beyond $3 \me$ could occur after dispersal of the gas disc.
Runs showing \emph{moderate growth and migration} formed planets in the mass range
$3 < m_{\rm p} < 35 \me$ during the gas disc life time. Simulations categorised as
\emph{giant formation and significant migration} formed planets with masses $\ge 35 \me$
during the gas disc life time, and generally displayed multiple bursts of planetary accretion
accompanied by large scale migration that ended up with one or more planets migrating 
into the central star. The planets that are formed in the simulations have different 
compositions in terms of rocky, icy and gaseous material. We use a classification system 
for the planets based on their compositions, and these are defined in Table~\ref{tab:plcompo}.

\subsection{Typical behaviour}
\subsubsection{Disc evolution with an active inner turbulent region}
\begin{figure*}
\includegraphics[scale=0.28]{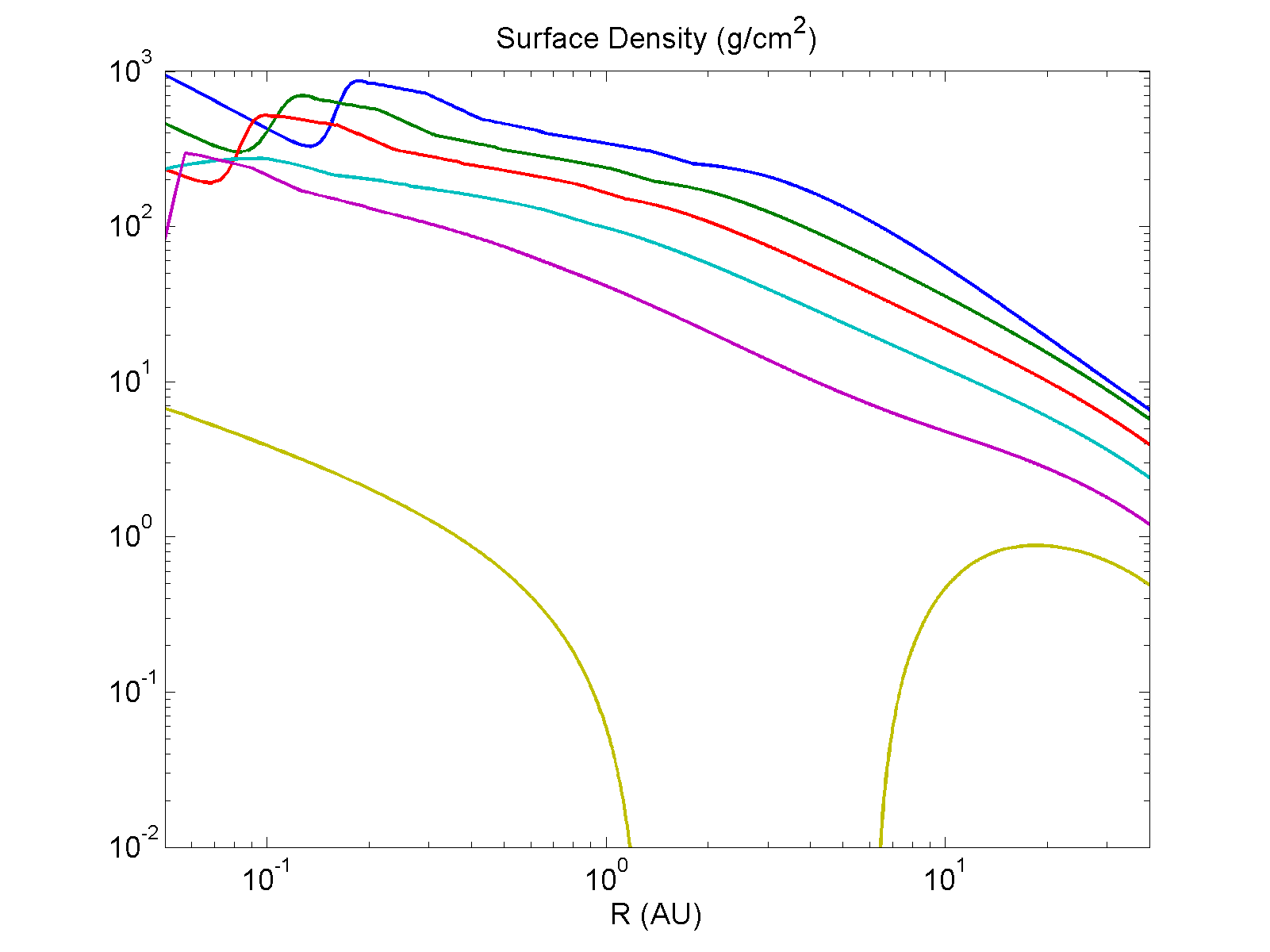}
\includegraphics[scale=0.28]{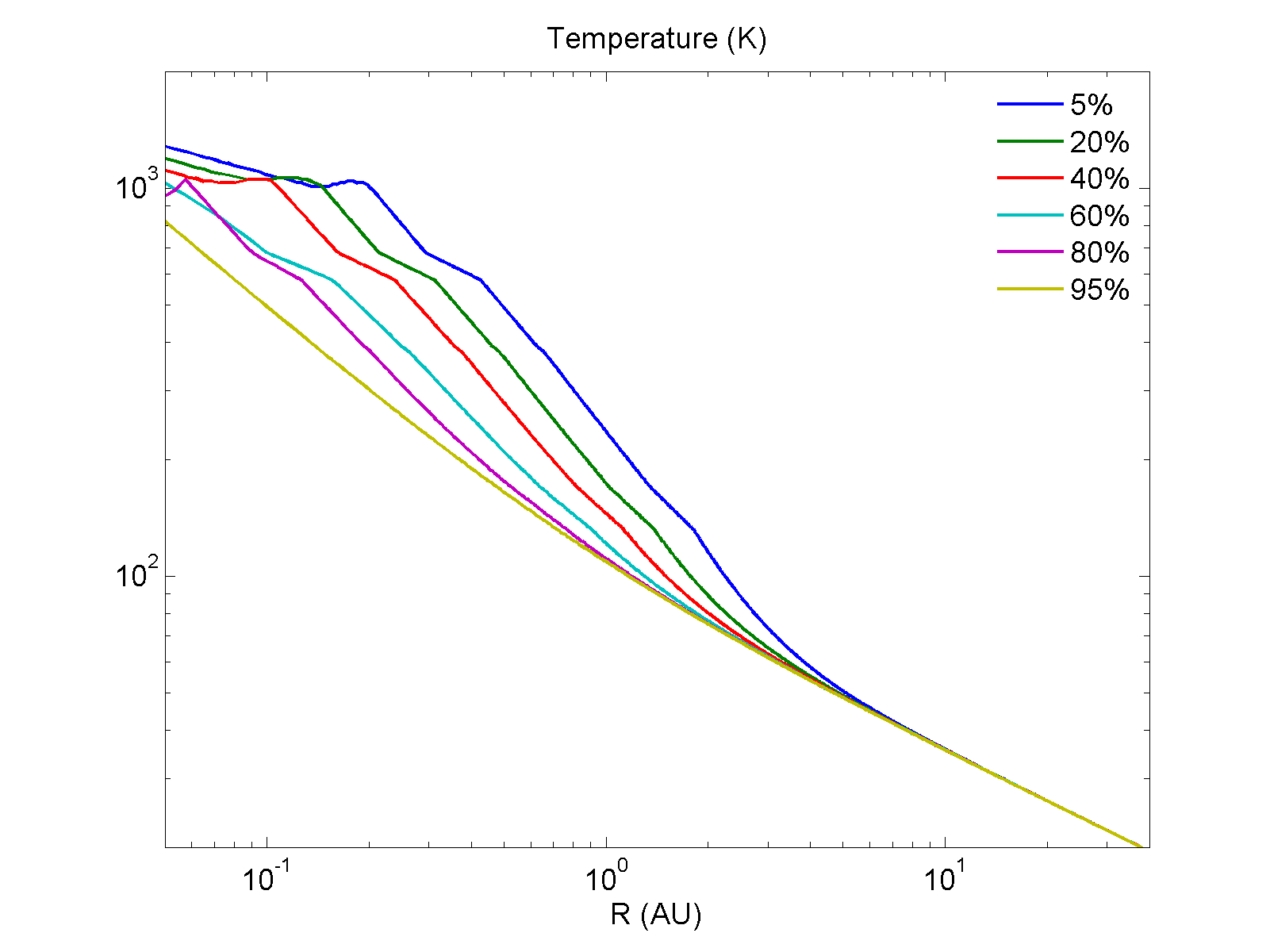}
\includegraphics[scale=0.28]{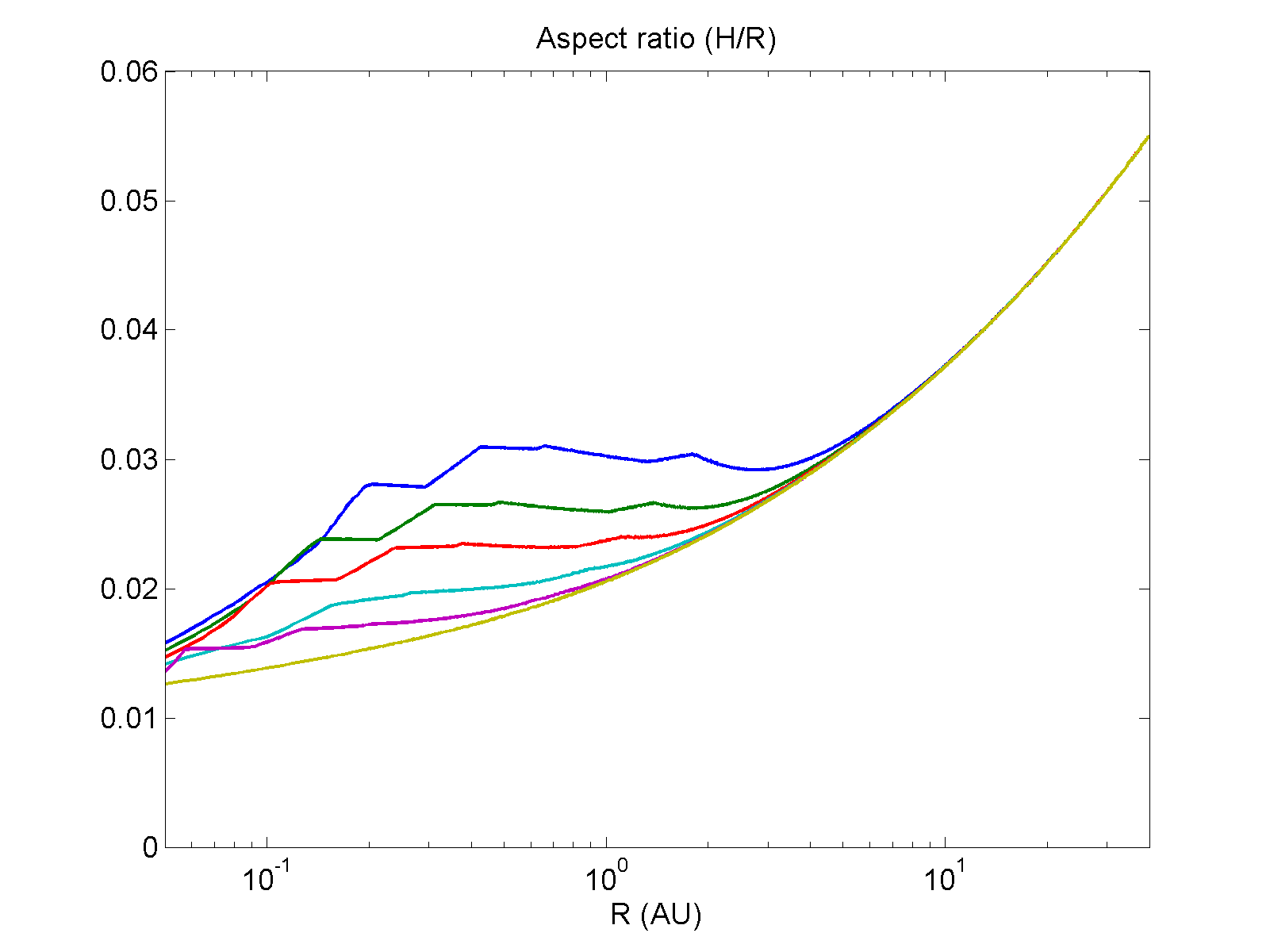}
\caption{Gas surface densities, temperatures and aspect ratios for 5, 20, 40, 60, 80, 95\% (top-bottom lines)
of the disc life time in a $1 \times \mmsn$ disc (life time: 4.6 Myr)}
\label{fig:multiplot}
\end{figure*}

Figure \ref{fig:multiplot} shows the evolution of a  $1 \times \mmsn$ disc model. Disc surface density profiles are shown in the left panel, temperature profiles 
are shown in the middle panel, and $H/r$ profiles are shown in the right panel. The times corresponding to each profile are indicated in the middle panel, expressed as a percentage of the disc lifetime. For a  $1 \times \mmsn$ disc this is equal to 4.6 Myr. For a $1.5 \times \mmsn$ disc the life time is 5.5 Myr, and a  $2 \times \mmsn$ disc disperses completely after 6.5 Myr. The inclusion of
a turbulent inner region where $T>1000$~K causes a dip in surface density due to 
the higher viscosity there, and it can be seen that as time progresses the location
of the transition to the turbulent region moves in towards the star because the
reduction in surface density reduces the viscous heating rate and the opacity.
The turbulent region disappears when the disc temperature no longer exceeds 1000~K
anywhere in the disc, as shown by the yellow line in Figure \ref{fig:multiplot}.
This happens in all of our disc models when the disc mass falls to approximately
10\% of the MMSN, which occurs 0.5 Myr before complete dispersal of the gas disc.

The drop in local surface density caused by the active turbulent region creates 
a planet trap for low-mass planets \citep{Masset2006}. 
The trap moves in with the active region until it reaches the inner disc edge 
located at $0.05 \au$
(assumed in our model to be outer edge of the magnetospheric cavity). 
\textbf{Once at the disc inner edge, the trap created from the active turbulent region disappears due to the temperature in the disc falling below 1000~K. However the outer edge of the magnetospheric cavity acts as a planet trap for low-mass planets, until they can open a gap in the disc and undergo type II migration into the cavity as discussed in Section \ref{sec:magnetospheric_cavity}.}
It should be noted that the reduction of the temperature below 1000~K at all disc locations arises
because of our adoption of a 1D disc model which neglects irradiation heating of the disc along radial lines of sight, as discussed in Section~\ref{sec:discussion}. 

On longer time scales the removal of gas by the photoevaporative wind causes the disc to disperse. The loss of mass at large radius results in the inner disc emptying viscously onto the star, followed by removal of the remnant outer disc by the wind
\citep{Clarke2001}.

\begin{figure}
\includegraphics[scale=0.45]{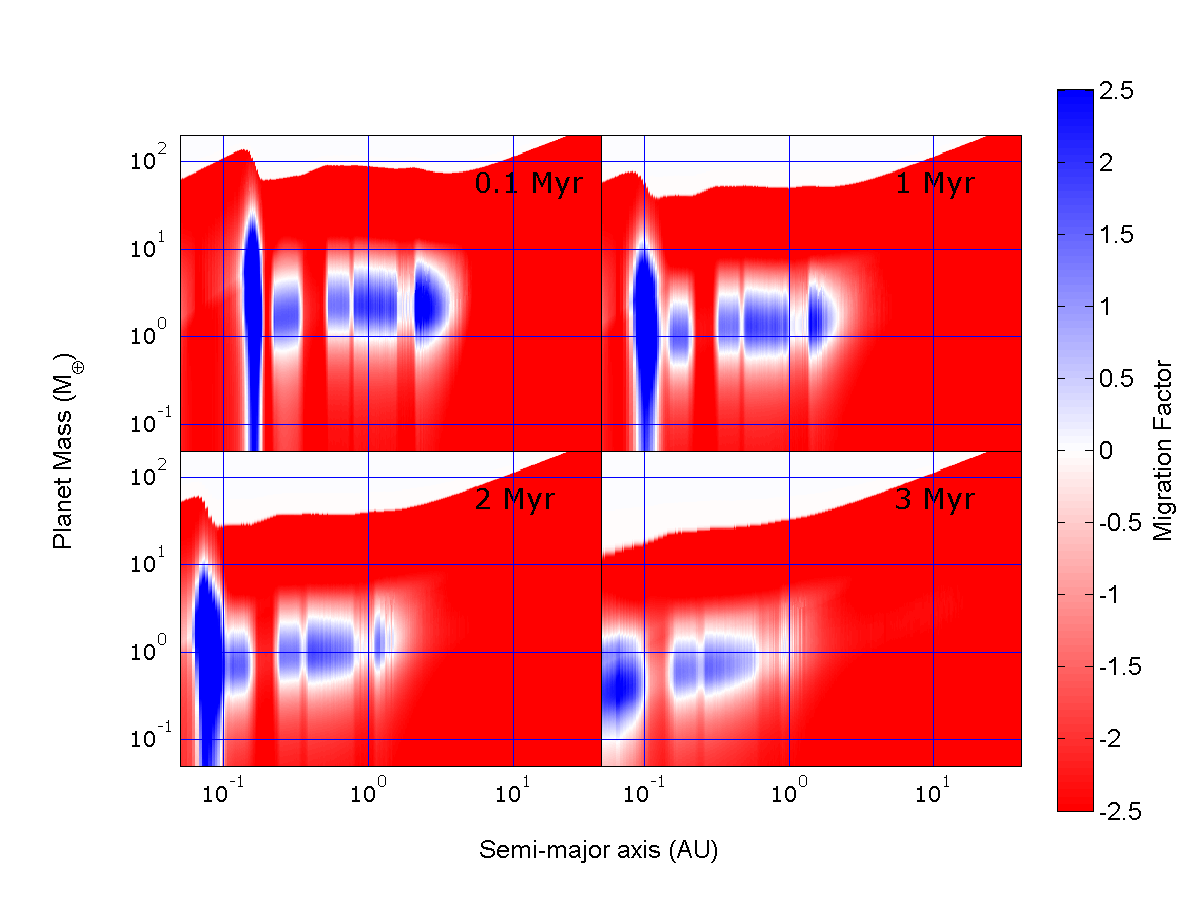}
\caption{Contour plots showing regions of inwards (red) and outwards (blue) migration) in a $1 \times \mmsn$ disc at t = 0.1 Myr (top left), t = 1 Myr (top right), t = 2 Myr (bottom left) and t = 3 Myr (bottom right).}
\label{fig:contours}
\end{figure}

\subsubsection{Protoplanet migration}
\label{sec:common_migration}
Type I migration of planets is controlled by both Lindblad and corotation
torques. In our disc models Lindblad torques are negative and corotation
torques are generally positive. Strong, positive corotation torques arise in regions where the 
radial entropy gradient is negative, and this is usually the case in the 
inner disc regions where viscous heating dominates over stellar irradiation. 
Corotation torques may saturate when either the viscous or thermal time scale 
differs significantly from the periods of horseshoe orbits executed by gas in
the corotation region. 
Figure~\ref{fig:contours} shows contours that illustrate the migration behaviour 
of planets as a function of their masses and semimajor axes in a $1 \times \mmsn$
disc with solar metallicity where half of the solid material is assumed to
be in large bodies that do not provide any opacity. Dark blue regions
correspond to strong outward migration, red regions correspond to strong
inward migration, and white contours represent regions of parameter
space where the corotation and Lindblad torques balance each other. 
We refer to these as \emph{zero-migration zones}. The planet trap created
by the inner turbulent region is shown by the innermost blue contour
in the first three panels in Figure~\ref{fig:contours}. Planets in blue
regions migrate outwards until they come to white regions where they stop
migrating. These can and do act as planet convergence zones. Planets in red regions
migrate inwards, and if their masses are in the appropriate range they stop
when they arrive at zero-migration zones. Over time we see that the migration
contours evolve as the disc surface density and thermal time scale decrease,
and planets sitting in zero-migration zones slowly drift in towards the star on 
the disc evolution time scale. A planet that grows in mass so that it exceeds 
$\sim 10 \me$ will be too massive to sit in a zero-migration zone in the main 
body of the disc, and will migrate inwards rapidly before being trapped at the 
transition to the inner turbulent region. As this disappears the planet will drift 
into the magnetospheric cavity interior to $0.05 \au$ where it will stop if it is
below the local gap forming mass. If it exceeds the gap forming mass then it will
migrate to the 2:1 resonance location with the cavity outer edge before halting its 
migration. If another planet enters the cavity then it may push the previous one
through the inner boundary of the computational domain interior to $0.02 \au$.
The decrease in $H/r$ values in the inner disc regions (and with time) means that
it becomes possible for quite low mass planets to open gaps in the disc and
enter type II migration. Similarly, planets that accrete significant
gas envelopes can become giant planets and open gaps. The transition to gap
formation and type II migration is shown by the boundary between the red and white 
contours in the top regions of the panels in Figure~\ref{fig:contours}.
\begin{figure*}
\begin{center}
\includegraphics[scale=0.35]{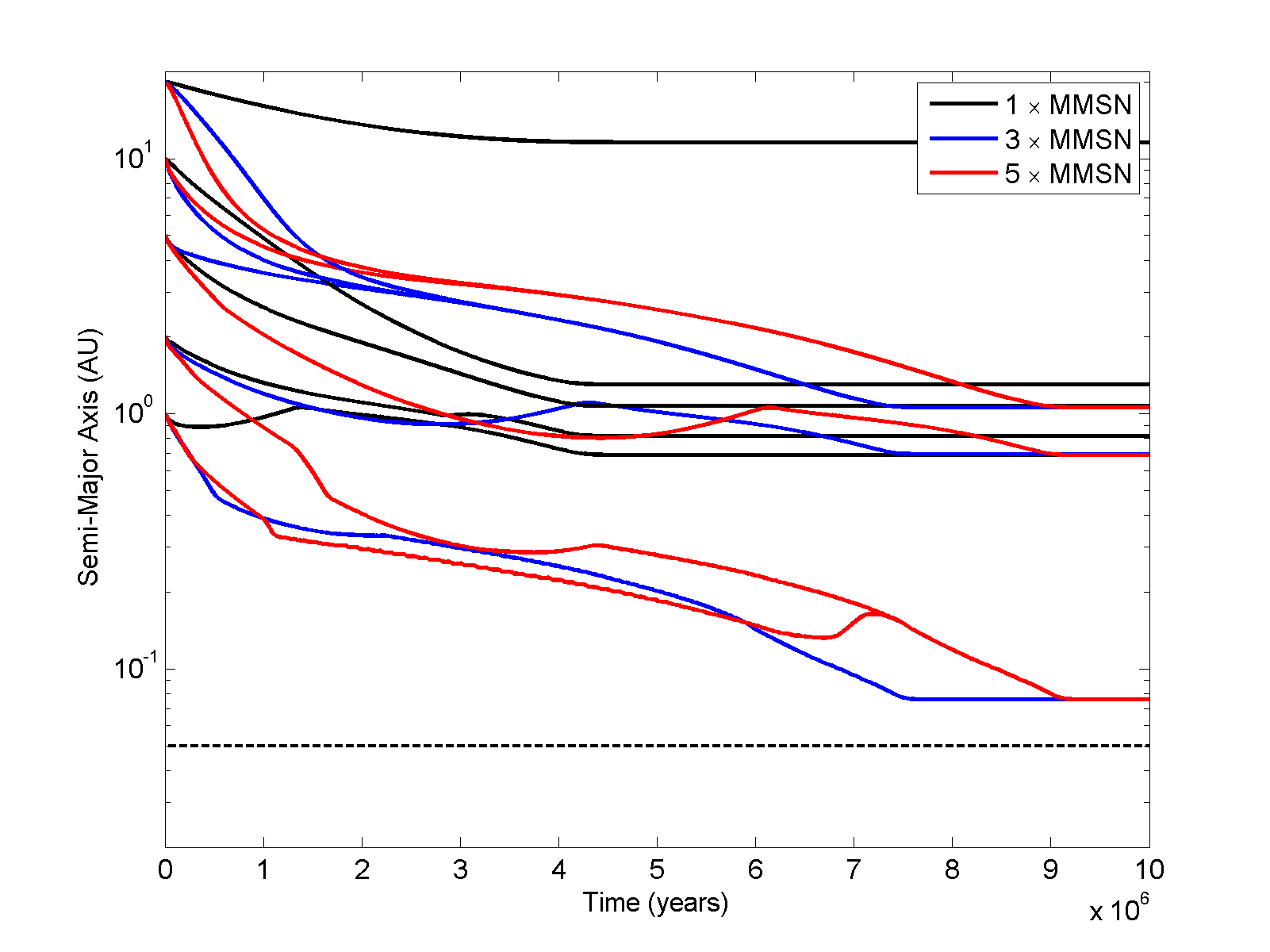}
\includegraphics[scale=0.35]{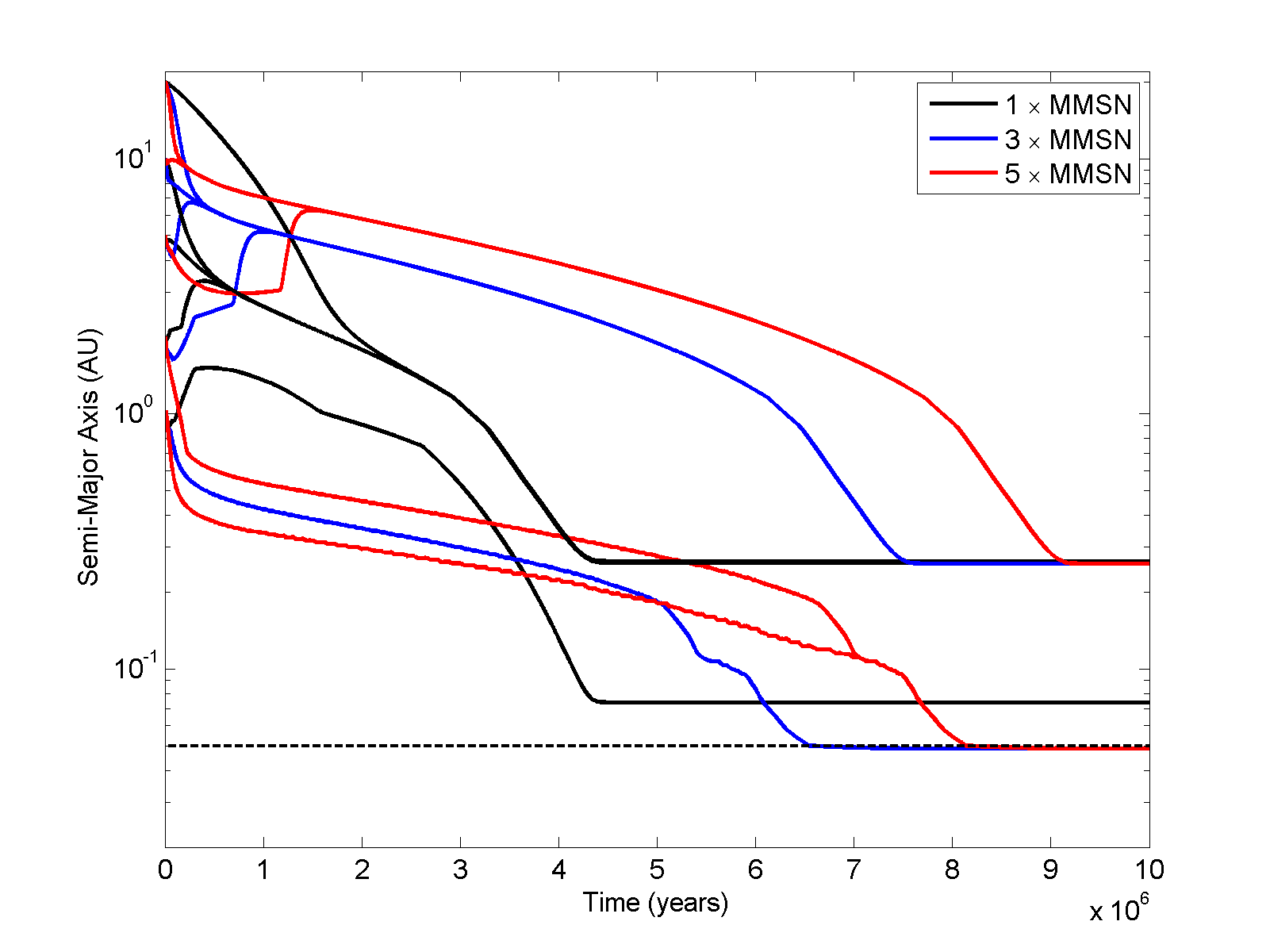}
\includegraphics[scale=0.35]{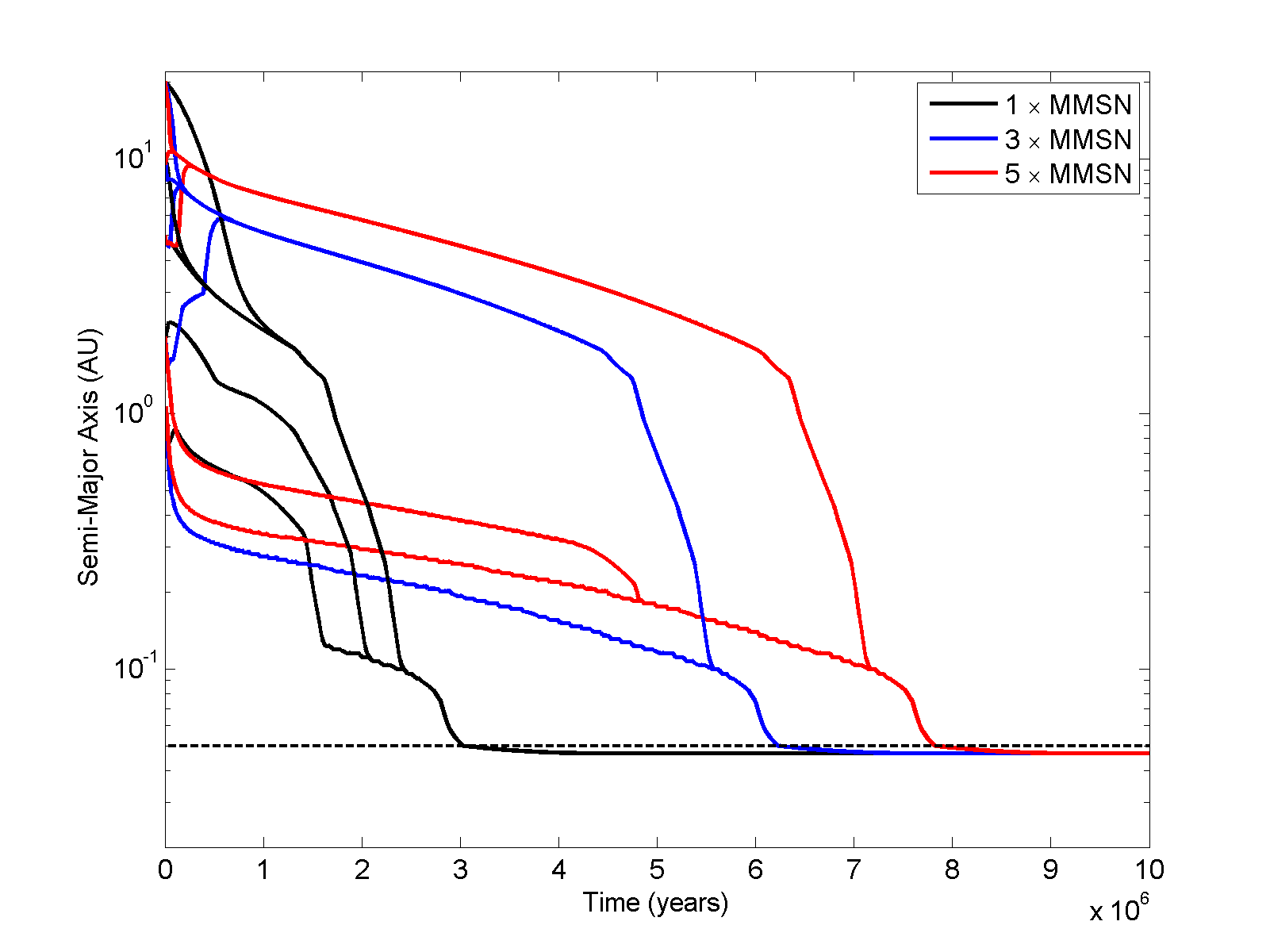}
\includegraphics[scale=0.35]{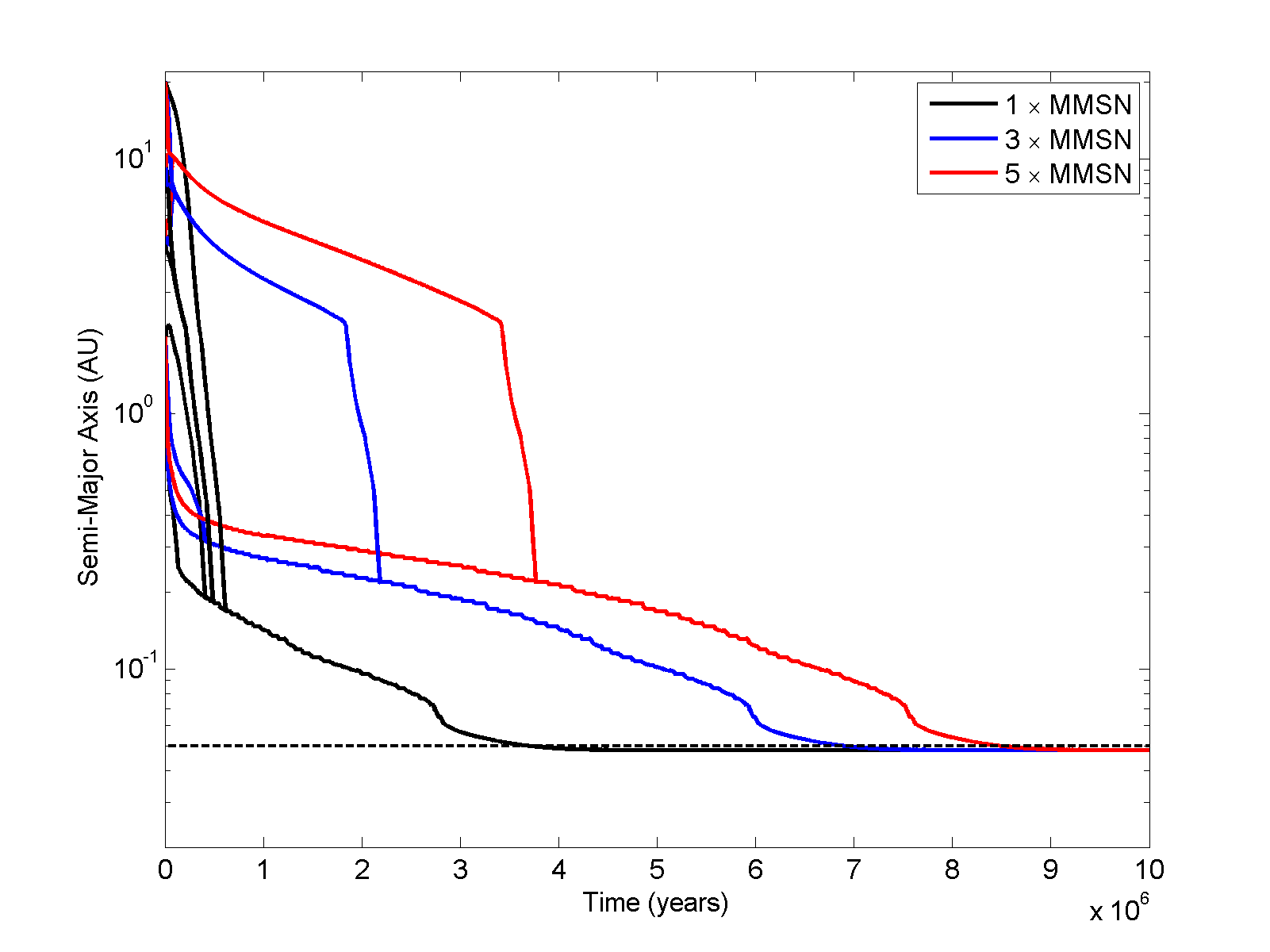}
\caption{Semimajor axis evolution for planets with different masses in 1, 3 and $5 \times \mmsn$ discs: $1 \me$ (top left), $3 \me$ (top right), $5 \me$ (bottom left) and $10 \me$ (bottom right). The dotted line represents the disc inner edge.}
\label{fig:planetevolution}
\end{center}
\end{figure*}

Each panel in Figure \ref{fig:planetevolution} shows the migration histories of individual 
planets of mass $1 \me$ (top left), $3 \me$ (top right), $5 \me$ (bottom left) and $10 \me$ (bottom right) 
embedded in discs with masses 1, 3 and $5 \times \mmsn$. In each panel we plot the migration tracks of planets that have initial semimajor axes 1, 2, 5, 10 and 20 $\au$.
Note that we only consider disc masses in the range 1 - 2 $\times$ MMSN in the full 
N-body simulations described below, but we include larger disc masses in this discussion 
to illustrate how migration changes in significantly heavier discs. Looking at the $1 \me$
migration trajectories it is clear that planets starting with $a_{\rm p} \ge 1 \au$ 
in a $1 \times \mmsn$ disc cannot migrate interior to $0.7 \au$ because of the corotation
torques. Even in heavier discs $1 \me$ planets cannot migrate very close to the star and
become stranded outside the magnetospheric cavity at $\sim 0.07 \au$. The implications of
this are clear. The origin of compact, short-period low mass planet systems such as Kepler
444 \citep{Campante2015} or Kepler 42 \citep{Muirhead2012} cannot be explained by formation 
at significantly larger radii than where they are observed today, followed by large scale
inward migration. An \emph{in situ} formation model, perhaps aided by the 
inward drift of solids in the form of pebbles, boulders or small planetesimals would seem to be 
more plausible. More generally, \emph{in situ} models of planet building cannot rely on the
delivery of large numbers of low mass protoplanets to inner disc regions through type I
migration because they are not able to migrate across the required distances during gas disc life 
times. Looking at the $3 \me$ migration trajectories, we see that these planets are also
unable to reach the inner magnetospheric cavity unless orbiting in heavier discs.
Guaranteed arrival of planets to the very innermost regions of the disc only occurs
when planet masses reach $m_{\rm p} \ge 5 \me$. Periods of rapid migration observed
in the lower left and right panels of Figure \ref{fig:planetevolution} arise when
the planets saturate their corotation torques. Slow drift arises when the planets
are sitting in zero-migration zones.

\subsubsection{Planetesimal orbital evolution}
Aerodynamic drag causes planetesimal eccentricities and inclinations to be damped
and their semimajor axes to decrease. The 10~m boulders in our simulations experience 
rapid migration such that a body located initially at $1 \au$ migrates to the 
inner turbulent region of the disc within aproximately $10^3$ years, and a 10~m boulder located
at $20 \au$ reaches there in just over $10^6$ years. A 100~m body located
initially at $1 \au$ reaches the inner turbulent region within $\sim 0.5$ Myr, 
and one located initially at $10 \au$ will reach $\sim 6 \au$ within the disc life time. The larger 1~km and 10~km
bodies show very little drag-induced migration during disc life times.

The levels of planetesimal/boulder eccentricity excitation due to gravitational stirring 
by protoplanets at the beginning of the simulations depends strongly on their sizes. 
We find that the mean eccentricity for the 10~m bodies is $e_{\rm pl} \sim 3$ - 
5 $\times 10^{-4}$, for the 100~m bodies $e_{\rm pl} \sim 3$ - 4 $\times 10^{-3}$, 
for the 1~km bodies $e_{\rm pl} \sim 10^{-2}$ and for the 10~km planetesimals 
$e_{\rm pl} \sim 2$ - 3 $\times 10^{-2}$.
Given the importance of gravitational focussing in determining planetary growth rates, it is clear
that we should expect smaller boulders/planetesimals to accrete much more efficiently onto the 
protoplanets. The mobility of the boulders also means that planetary embryos can grow beyond their
nominal isolation masses on short time scales before they start to undergo significant type I migration. For protoplanets whose masses are too small for type I migration,
it is the mobility of boulders and small planetesimals in our models that enables growth 
to occur above the isolation mass.

\begin{figure*}
\includegraphics[scale=0.8]{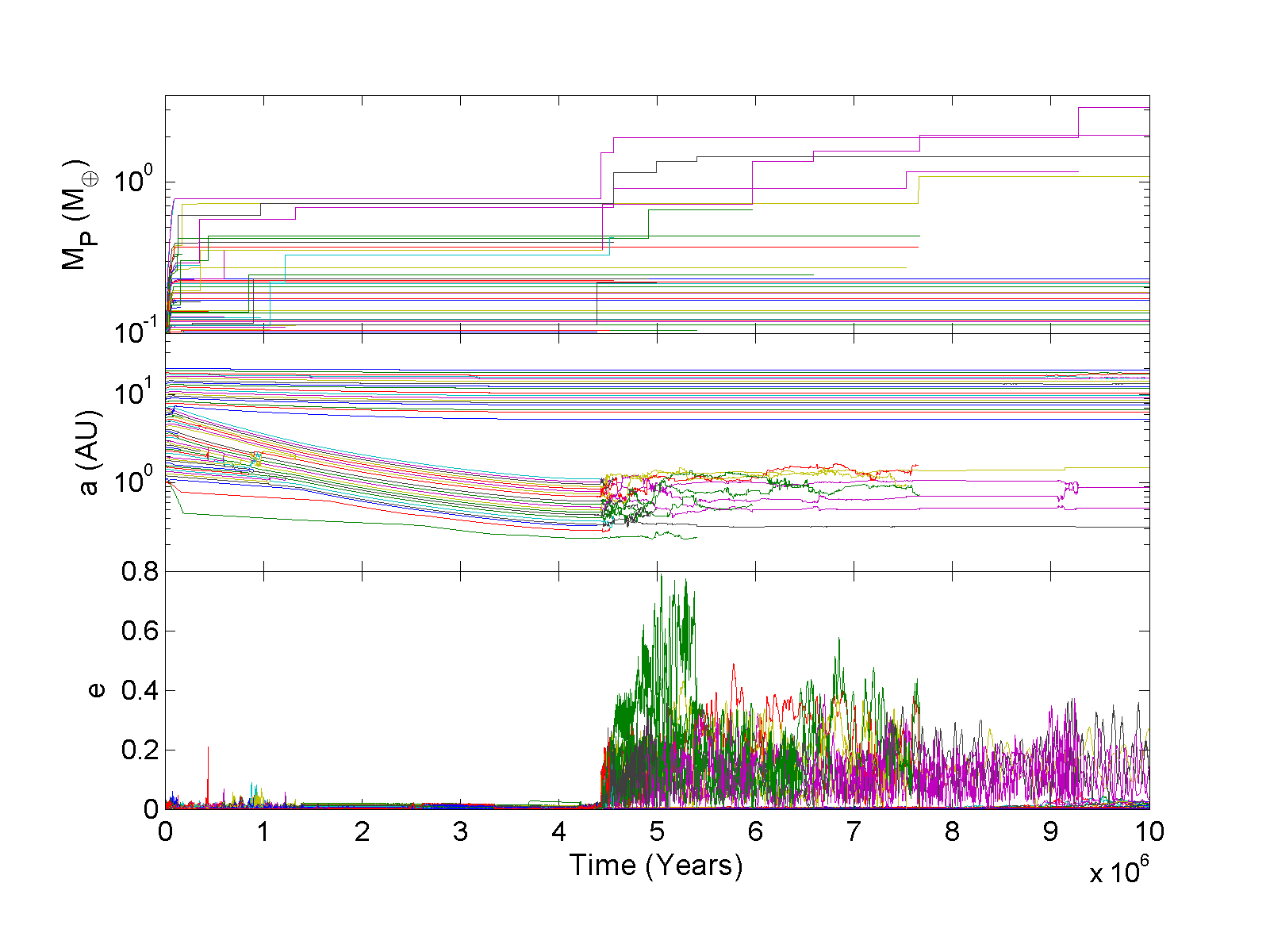}
\caption{Evolution of masses, semi-major axes and eccentricities of all protoplanets 
in simulation K10.50.01B.}
\label{fig:K10.50.01Bmulti}
\end{figure*}

\subsubsection{Recap on results from CN14}
The simulations presented in CN14 adopted an inner boundary to the computational
domain at an orbital period of 20 days, and so those calculations were unable to 
explore the formation of short period compact planetary systems. Disc masses between 
$1 \times$ and $5 \times \mmsn$ were considered, with metallicities of $1 \times$ and  $2 \times$ solar.
Planetesimal sizes were 1~km and 10~km. Unlike in the present runs, no reduction in 
opacity was imposed to account for the growth of submicron dust grains into 
planetesimals/boulders and planetary embryos that populate the disc at the beginning
of the simulations. Instead, the inconsistent initial condition that all disc solids 
were in the form of planetesimals and protoplanets, but with no diminution of the opacity, 
was adopted. The disc model with $1 \times \mmsn$ and solar metallity in CN14 is 
therefore equivalent to the model with $1 \times \mmsn$ and $2 \times$ solar metallicity 
in this paper. The main results of CN14 can be summarised as:
\begin{itemize}
\item For discs with a low to moderate abundance of solids, only limited growth of
planets was observed before gas disc dispersal, although growth of planets to masses
$m_{\rm p} \sim 6 \me$ was observed due to continued mutual collisions after the disc
was gone. As a consequence, only modest migration was observed in these runs, such that
essentially no material was lost through the inner boundary.
\item It was commonly observed that numerous super-Earths and Neptune-mass planets
formed in discs with intermediate masses. The bodies frequently migrated out of the disc
through the inner boundary, such that in some runs no planets were left in the system at all.
In others, a few super-Earth and Neptune-mass planets were able to survive.
\item The highest disc masses considered usually led to multiple bursts of planets forming.
Gas giants with masses $m_{\rm p} > 35 \me$ formed frequently. In all cases, these planets 
migrated rapidly through the disc via type I and II migration and out through the inner boundary.
In some runs the final burst of planet formation and migration led to the formation of
a short-period compact system of super-Earths and Neptunes that was able to survive.
The highest mass planet to survive in all runs was a $13 \me$ gas-rich Neptune.
\item CN14 determined the conditions under which giant planets could form and avoid migration
into the star in their model of planet formation. They showed that the disc mass and orbital 
radius at which a core starts to undergo runaway gas accretion and type II migration need to be 
$\sim 0.6$ $\times$ MMSN and $a_{\rm p} \gtrsim 8 \au$, respectively. These same conditions 
also apply to the models that we present in this paper, except that our adoption of a
magnetospheric cavity prevents an individual planet migrating all the way out of the computational
domain interior to $0.02 \au$. It is extremely difficult in our disc model for a giant planet
core to form and undergo runaway gas accretion at orbital radii $> 8 \au$ because the
zero-migration zones shown by the contours in Figure~\ref{fig:contours} migrate in too quickly
to favour such an outcome. We note that increasing the disc viscosity and changing
assumptions about the opacity model can favour the trapping of higher mass
planets in zero-migration zones at larger orbital radii early during disc life times, but these zero-migration zones drift downwards and inwards more rapidly in
the mass-period plane (as plotted in Figure~\ref{fig:contours}) in these models because 
of the faster disc evolution \citep[e.g.][]{Bitsch}. 
\end{itemize}

\begin{figure*}
\includegraphics[scale=0.8]{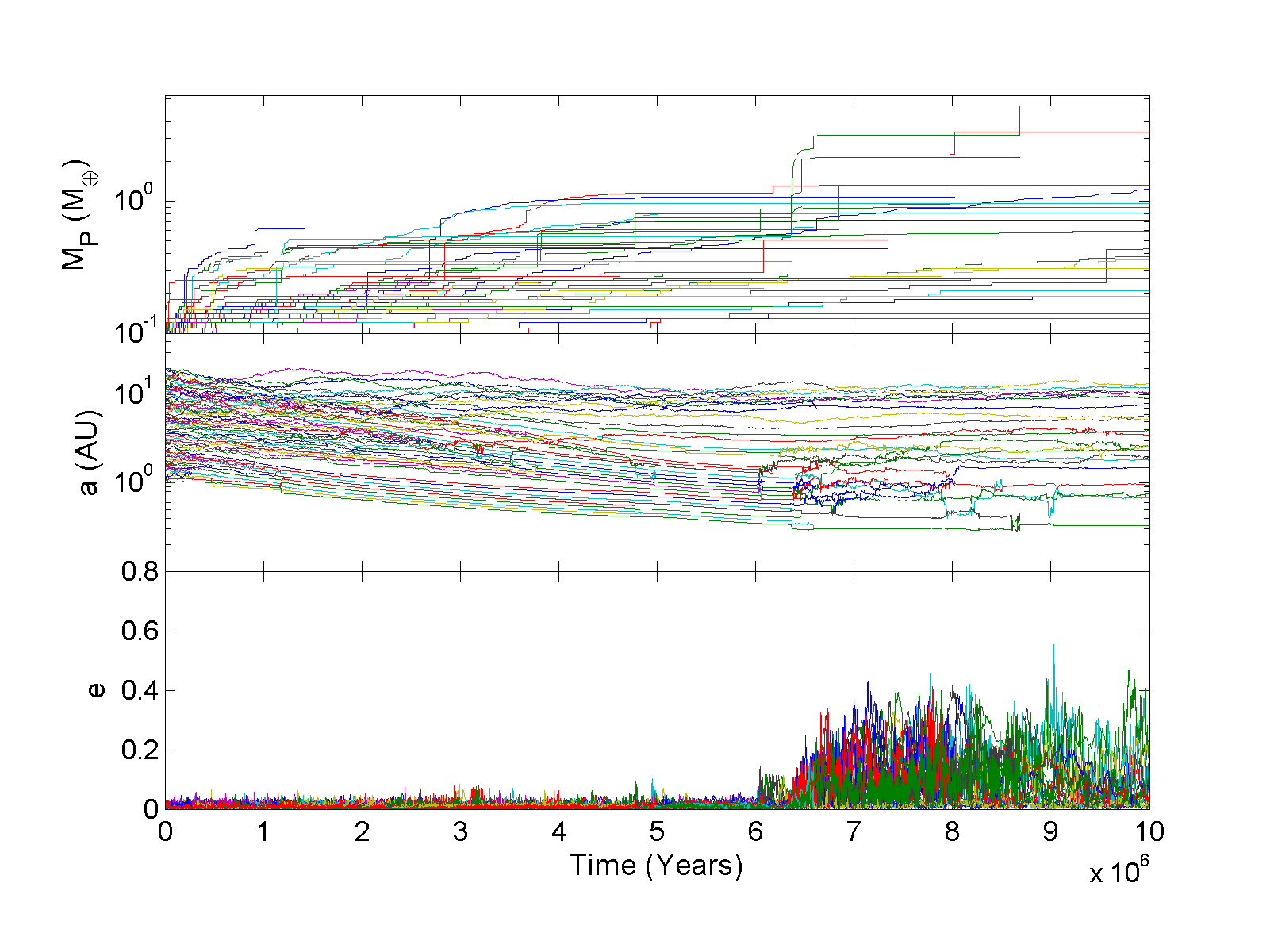}
\caption{Evolution of masses, semi-major axes and eccentricities of all protoplanets 
in simulation K2210B.}
\label{fig:K2210Bmulti}
\end{figure*}

\subsection{Limited Planetary Growth (LPG)}
\label{sec:limited_growth}
The mass growth of planets is expected to be slow when either the 
abundance of solids in the disc is small, and/or when the main feedstock for planet
building is in the form of large planetesimals whose velocity dispersion is damped weakly by the gas disc. Consequently, in the limit of slow growth, no gas accreting 
cores with masses $m_{\rm p} \ge 3 \me$ will be able to form before dispersal of the gas 
disc\footnote{We note that planetary atmospheres may form {\it via} outgassing, but 
this effects goes beyond the range of physical processes considered in our models. 
Furthermore, H/He rich envelopes can settle onto relatively low mass planets 
\citep{Lammer}, and although we consider the effect of this on planetesimal accretion, 
we do not report gas envelope masses for planets with $m_{\rm p} < 3 \me$ in this paper.}, 
and planet migration will be modest.
This outcome was obtained for all but one disc model that we considered with planetesimal 
sizes being either 1 or 10~km (the exception being the heaviest disc with mass 2 $\times$ MMSN, $2 \times$ solar metallicity and 1~km planetesimals). At the other end of the boulder/planetesimal size scale when 10~m boulders were included in the runs, this outcome was obtained only for the disc model with the lowest mass and metallicity.
Overall, these results are in agreement with the low solid abundance models presented in CN14.

The simulations labelled as LPG in Table \ref{tab:simparam} all displayed this mode of behaviour, 
and below we describe in detail the results of runs K10.50.01B and K2210B as they have
very different disc properties, but result in similar outcomes.

\begin{figure*}
\includegraphics[scale=0.8]{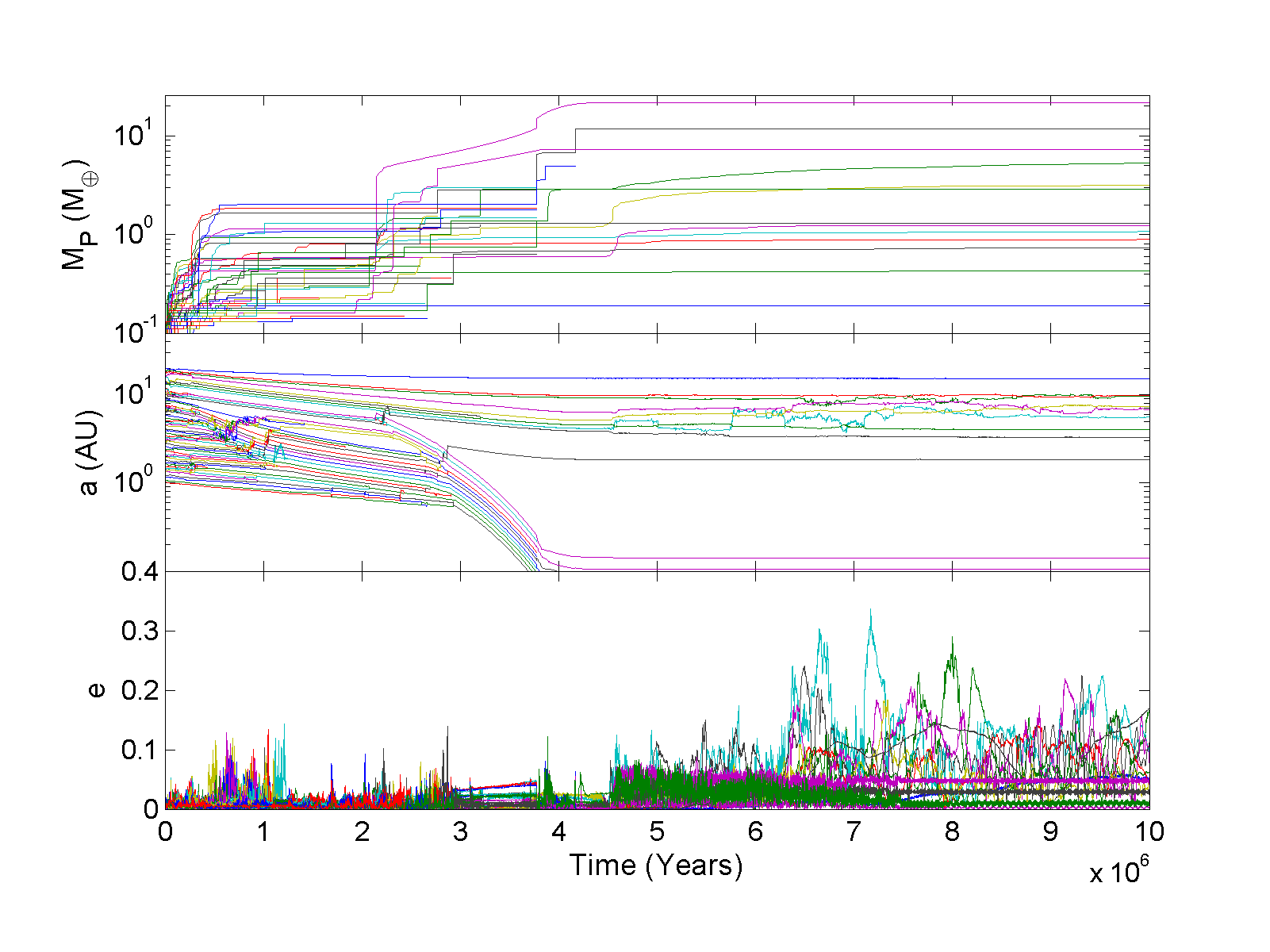}
\caption{Evolution of masses, semi-major axes and eccentricities of all protoplanets 
in simulation K120.1B.}
\label{fig:K120.1Bmulti}
\end{figure*}

\subsubsection{Run K10.50.01B}
Run K10.50.01B had a disc mass of $1 \times \mmsn$, $0.5 \times$ solar metallicity, 
and boulder radii $R_{\rm pl}=10$~m. The combined mass in protoplanets and 
boulders was equal to $11 \me$, distributed between $0.5 \le r \le 20 \au$, with 
the mass in protoplanets being initially $5.2 \me$ (52 protoplanets each of
mass $0.1 \me$).

The evolution of the protoplanet masses, semimajor axes and eccentricities are shown in 
Figure~\ref{fig:K10.50.01Bmulti} (note that boulders/planetesimals are not represented in this
and similar plots).
Accretion of boulders by embryos, and mutual collisions, led to the 
growth of protoplanets to masses in the range $0.6 \le m_{\rm p} \le 0.8 \me$ during
the first 1~Myr. These embryos migrated towards the zero-migration zone located at 
$\sim 3 \au$ and drifted in towards the star on the disc evolution time. Embryos located 
beyond $10 \au$ grew more slowly, and remained near their initial locations throughout the 
simulation. We note that a couple of embryos at the inner edge of the solids disc experienced 
a short lived burst of migration by being shepherded inwards by a swarm of migrating boulders 
at the beginning of the simulation.

Despite the convergence of planets in the zero-migration zone, the frequency of collisions 
was limited by bodies entering mean motion resonances. Boulder collisions with embryos were scarce after 1~Myr, due to the drag-induced migration of boulders into the inner disc occurring on this time scale. With the maximum mass of a planetary embryo in the system 
being $0.8 \me$ throughout the life time of the gas disc, migration remained limited to a slow inwards drift. No planets accreted gaseous envelopes.

The disc photoevaporated after 4.6~Myr, allowing embryo eccentricities to grow 
dramatically through mutual encounters because gas disc damping had been removed.
Collisions among the inner group of protoplanets led eventually to the formation
of a system of four inner bodies with masses in the range $1.1 \le m_{\rm p} \le 3.4 \me$
after 10~Myr when the simulation ended. These bodies all accreted significant amounts of 
material from beyond the snowline, and we class them as either water-rich terrestrials 
or water-rich super-Earths, orbiting with periods $60 \le  P \le 700$ days.
There were a significant number of protoplanets orbiting exterior to $5 \au$ still undergoing 
collisional evolution at 10~Myr when the simulation ended, and these would have continued accreting if the run had been extended.

\subsubsection{Run K2210B}
We turn now to run K2210B, for which the disc mass was $2\times \mmsn$, the metallicity was 
$2\times {\rm solar}$, planetesimal radii were 10~km, and the disc life time was 6.6~Myr.
The initial mass in embryos and planetesimals was $87 \me$, this being the most solids-rich disc considered in 
this paper. In spite of this, planetary growth was very limited because of the weakly-damped 
planetesimals.

The evolution of protoplanet masses, semimajor axes and eccentricities are 
shown in Figure~\ref{fig:K2210Bmulti}.
Protoplanets grew to masses $0.7 \me$ after 1 Myr, and when the disc dispersed
the maximum embryo mass was approximately $2.5 \me$, there having been a couple of planets that accreted rapidly just prior to the final remnants of the gas being removed. Migration was limited, with the
innermost body orbiting at $0.4 \au$ at the point of gas disc dispersal.
After removal of the gas the system entered a stage of chaotic evolution, with on-going
collisions occurring within the embryo swarm when the run ended at 10~Myr. Approximately
20 planets remained at this stage, the most massive being $m_{\rm p} = 5.3 \me$.
No planets accreted gaseous envelopes.

\subsection{Moderate growth and migration (MGM)}
\label{sec:moderate}
Table~\ref{tab:simparam} shows that a total of 16 out of 72 simulations exhibited \emph{moderate growth and migration}. MGM runs are characterised by the formation of planets with masses $3 \le m_{\rm p} < 35 \me$ before the end of the gas disc life time, with little or no loss of planets through the inner boundary of the computational domain. These simulations can result in two distinct planetary system architectures. One in which a dominant Neptune-mass body forms and migrates all the way into the magnetospheric cavity, and another where growth and migration of planets is more moderate, resulting in super-Earths and Neptunes orbiting at greater distances from the central star. Giants do not form because the growth of planets is slow enough that gas envelope accretion starts late during the disc life time, such that only moderate envelope masses have time to accrete. We discuss one representative
example of an MGM run that led to the formation of a compact system of super-Earths and Neptunes on short period orbits, but no planet orbiting within the magnetospheric cavity.

\begin{figure*}
\includegraphics[scale=0.8]{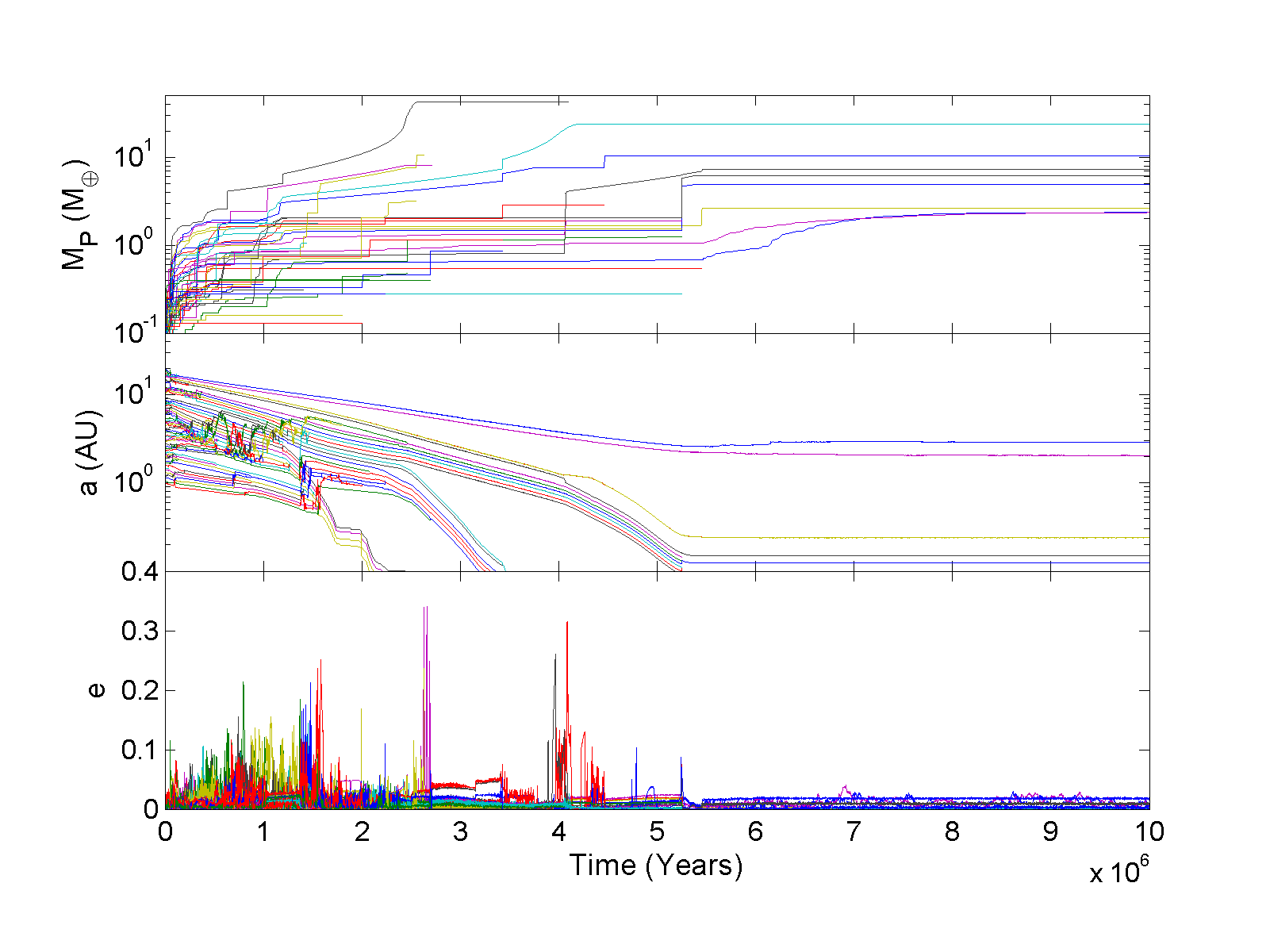}
\caption{Evolution of masses, semi-major axes and eccentricities of all protoplanets 
in simulation K1.520.1A.}
\label{fig:K1.520.1Amulti}
\end{figure*}

\subsubsection{Run K120.1B}
Run K120.1B had a disc mass of $1 \times \mmsn$, $2 \times$ solar metallicity,
and 100~m planetesimals. The total amount of mass in embryos and planetesimals was $43.5 \me$.

The evolution of embryo masses, semimajor axes, and eccentricities are shown in Figure \ref{fig:K120.1Bmulti}.
Several planets grew to masses $m_{\rm p} \sim 2 \me$ during
the first 0.5~Myr. A common phenomenon during the simulations involving 10~m boulders or 100~m planetesimals was the formation of shepherded rings of boulders/planetesimals while the gas disc was present, and from time to time rapid growth of a planet was observed if it crossed one of these rings through
embryo-embryo scattering. At 2~Myr an embryo of mass $0.43 \me$ located at $5.8 \au$ grew to $3.8 \me$ by accreting planetesimals from a shepherded ring, and hence started to accrete a gas envelope. The increase in mass eventually caused the corotation torques to saturate and the planet migrated in towards the star
before forming a gap and transitioning to slower type II migration
at $\sim 4$~Myr. Figure~\ref{fig:K120.1Bmulti} shows that the inward migration of this planet created an inward-migrating resonant convoy, with collisions between embryos and with planetesimals leading to embryos growing within the convoy. Initially consisting of 12 protoplanets, the arrival of
the convoy to the inner disc was followed by dynamical instability and collisions that left 4 short period planets remaining at the end of the simulation. These consisted of (moving out from the star) a $2.9 \me$ rocky terrestrial planet, an $11.6 \me$ gas-poor Neptune, a $7.2 \me$ mini-Neptune, and a $21.4 \me$ gas-rich Neptune, with orbital periods of 4.7, 8.3, 12.4 and 19.5 days, respectively.
As all of the orbital periods are less than 100 days, this inner group constitutes a compact system, within which
only one resonant pair exists, that being a 3:2 resonance between the gas-poor Neptune, and its neighbouring mini-Neptune.
Other resonances existed in this group of planets and their progenitors, but were broken when strong interactions and collisions occurred. This run provides a clear example of how a short-period compact system can form through concurrent growth and migration of planets.

In the outer disc regions beyond $2 \au$, the dispersal of the gas disc after 4.6 Myr led to dynamical excitation of the embryos orbiting there. Planetesimals rings that had been shephered by the planets were disrupted, and a number of planets grew in mass by accreting these planetesimals. At the end of the simulation the outer region was still undergoing active accretion, and would have led  eventually to the formation of long period water-rich terrestrial and super-Earth planets orbiting between $1.85 \le r_{\rm p} \le 15.2 \au$ if the run had been continued.

\subsection{Giant formation and significant migration (GFSM)}
\label{sec:colossal}
Table~\ref{tab:simparam} shows that only simulations with either 10~m boulders or 100~m planetesimals formed giant planets with masses $m_{\rm p} > 35 \me$. Out of 72 runs, 14 resulted in the formation of giants.

\begin{figure*}
\includegraphics[scale=0.28]{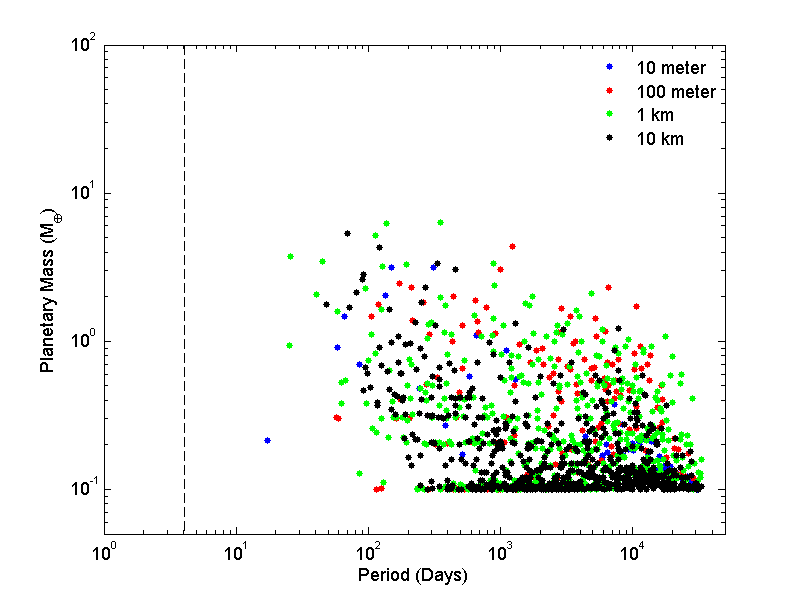}
\includegraphics[scale=0.28]{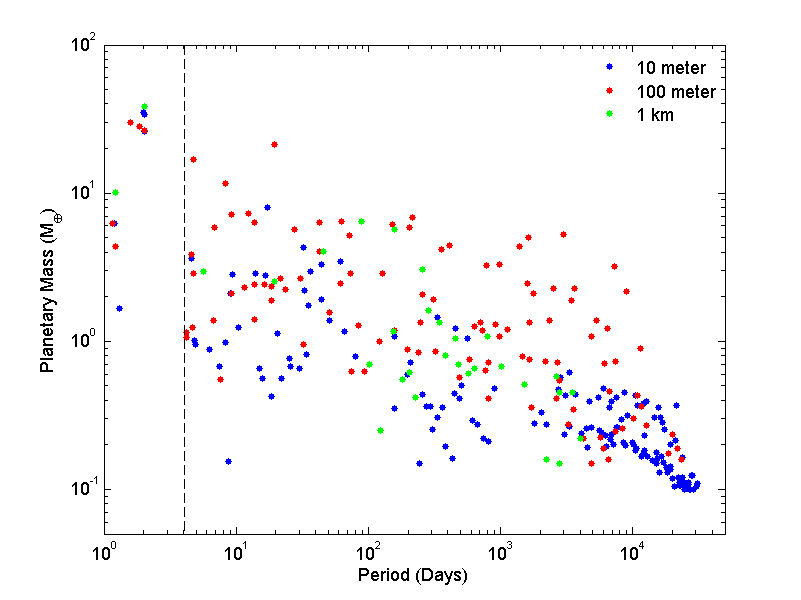}
\includegraphics[scale=0.28]{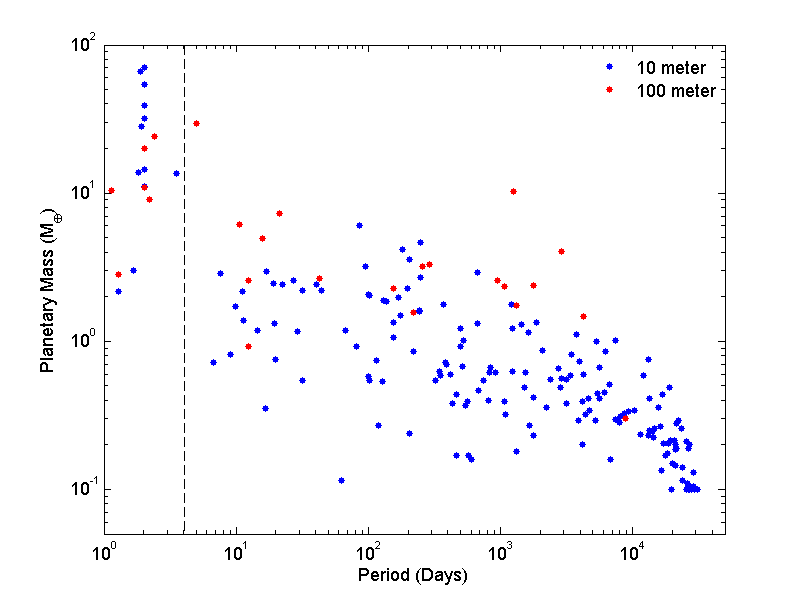}
\caption{Final masses versus orbital period for all planets formed in all simulations displaying \emph{limited planetary growth} (left panel), \emph{moderate growth and 
migration} (middle panel) and \emph{giant formation and significant migration} (right panel). 
Note that the runs are colour coded according to the 
planetesimal/boulder size adopted, as indicated in the legend 
in each panel.}
\label{fig:MVP-individual-runs}
\end{figure*}

Gas giant planet formation ensues because a core with 
$m_{\rm p} > 3 \me$ forms early enough that a substantial gas envelope can 
accrete either before the disc disperses or before the planet migrates into the inner magnetospheric cavity. In agreement with the results of CN14, we find that discs capable of forming giant planets undergo multiple bursts of planet formation and migration, with the first generation of giants being lost through the inner boundary. Unlike CN14, however, our model allows for the survival of migrating giants because they can become stranded within the magnetospheric cavity. Indeed, 
we formed a total of 5 surviving giants in the simulations, the most massive of 
which had $m_{\rm p} = 70 \me$. The most massive planet formed in any simulation
had $m_{\rm p} =160 \me$ (in model K220.01A), but was lost through the inner boundary because a second generation of planets arrived in the magnetospheric cavity and pushed it through the inner boundary interior to $0.02 \au$. We discuss one run below that formed giant planets that
experienced significant migration.

\subsubsection{Run K1.520.1A}
Simulation K1.520.1A had an initial disc mass of $1.5 \times \mmsn$, a solid abundance equal to 2 $\times$ solar and planetesimal radii 100~m. The mass in embryos and planetesimals was $65 \me$.

The evolution of protoplanet semimajor axes, masses and eccentricities are shown in Figure \ref{fig:K1.520.1Amulti}.
Two planets grew above $3 \me$ and started accreting gas envelopes within the 
first Myr. The saturation of corotation torques for the most rapidly growing
protoplanet caused it to migrate inwards, creating a resonant convoy of 
co-migrating interior embryos, one of which also accreted gas. The largest mass 
body that drove the migration of the chain reached $m_{\rm p} = 40 \me$ 
(with an envelope fraction of 87\%) before the convoy entered the magnetospheric
cavity. Gap formation prevented the $40 \me$ planet from halting at the
transition to the turbulent inner disc. The interior members of the group were pushed through the inner cavity and out of the computational domain, and the outermost planet stopped accreting gas and parked at the location of the 
2:1 orbital commensurability with the outer edge of the cavity.

Shortly after 1 Myr another pair of planets exceeded $3 \me$,
accreted gas envelopes and started to migrate rapidly
when their corotation torques saturated, driving another resonant
convoy inwards. These planets halted when they arrived at the transition to the active turbulent region at approximately 3.4~Myr. The outer planet in the convoy grew to $24 \me$, formed a gap and underwent type II migration into the magnetospheric cavity, pushing the resonant convoy and the earlier formed $40 \me$ giant
planet ahead of it. All the interior planets apart from an
adjacent $10.5 \me$ (formed by a collision within the cavity) were pushed through the inner boundary, leaving the $24 \me$ and $10.5 \me$ gas-rich Neptunes orbiting at 0.035 and $0.021 \au$ at the end of the simulation, with gas envelope fractions of 77\% and 32\%, respectively.

In the interval between 2 and 4 Myr a group of $\sim$ Earth-mass protoplanets
drifted in towards the star while sitting in a zero-migration zone, and halted their migration when the gas disc dispersed.
Subsequent collisions resulted in the formation of two water-rich super-Earths, a mini-Neptune and a water-rich terrestrial planet orbiting between $0.09 \au$ 
and $0.24 \au$ with masses in the range
$2.3 \le m_{\rm p} \le 8 \me$. At large radii (2 and $3 \au$, respectively) two water-rich terrestial planets are formed by the accretion of plantesimals after gas disc dispersal, reaching masses $\sim 2.5 \me$ at the end of the simulation at 10 Myr.

\subsection{Summary of LPG, MGM and GFSM results}
We now summarise the results obtained in the simulations
according to which class of outcome they fall into.

\subsubsection{LPG}
Simulations classified as showing 
\emph{limited planetary growth} led to similar outcomes
despite diverse initial conditions: 
(i) discs with low solids abundances containing boulders and small planetesimals; (ii) discs with relatively high abundances of solids in the form of large planetesimals.
The final outcomes of these simulations are summarised 
in the mass versus period diagram shown in the left panel of Figure~\ref{fig:MVP-individual-runs}. We see that no very
short period planets were formed, and final masses are all
below $10 \me$. The inverse correlation between mass and
semimajor axis arises because of modest disc driven
migration that caused the most massive bodies to drift in.
The colour coding of the symbols shows that the final 
outcomes are similar for all boulder and planetesimal 
sizes. 

\begin{figure*}
\includegraphics[scale=0.28]{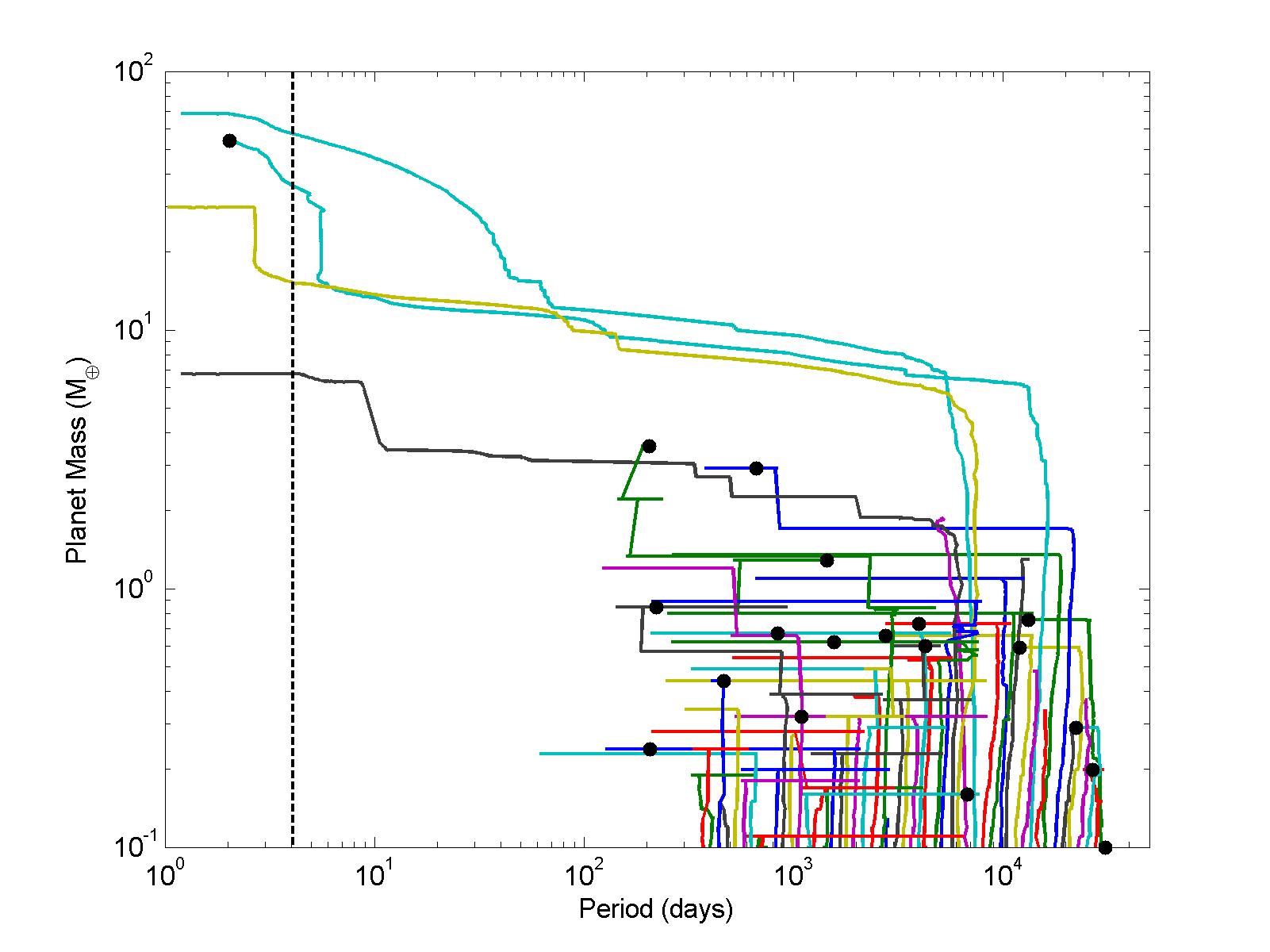}
\includegraphics[scale=0.28]{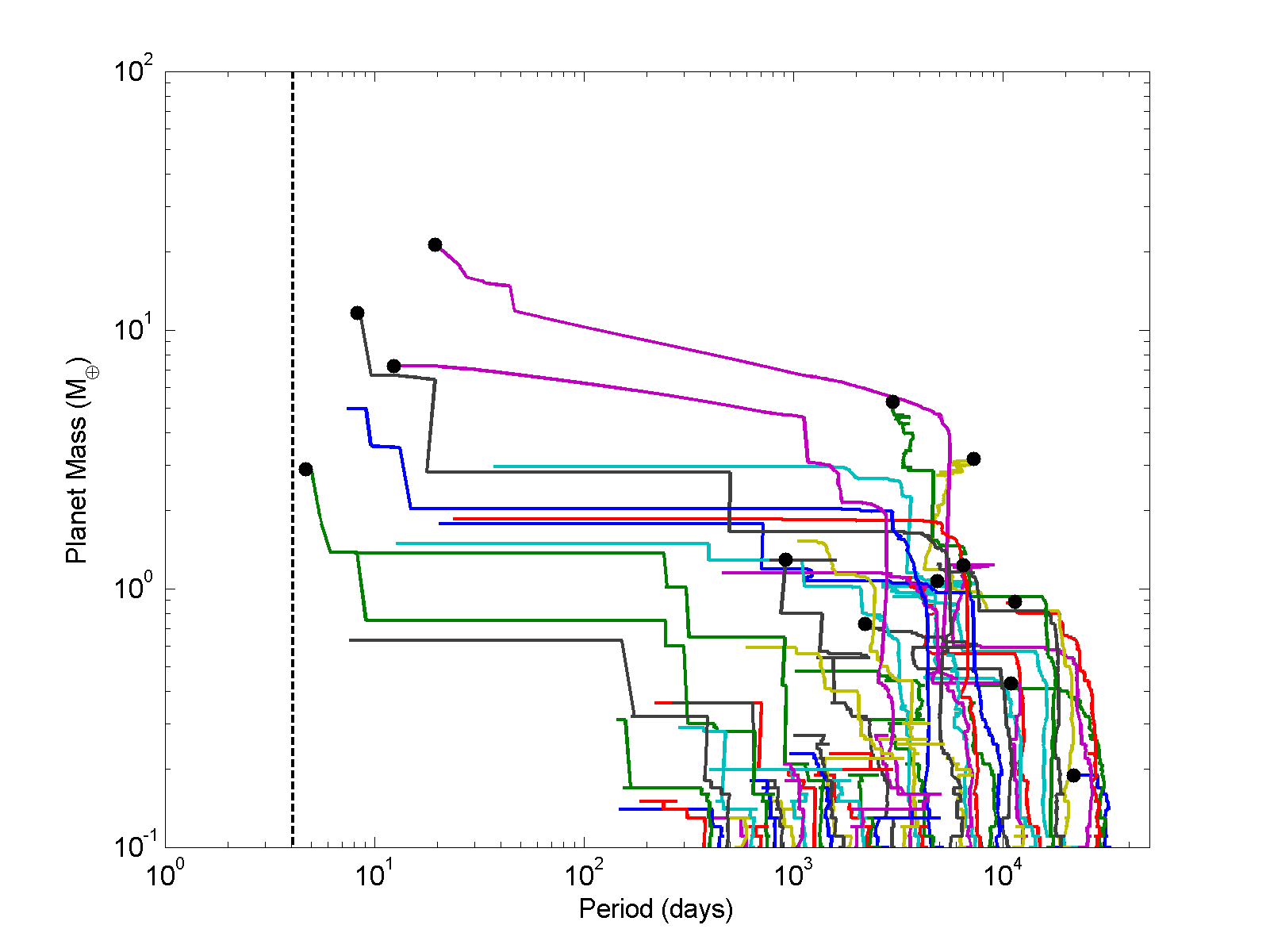}
\includegraphics[scale=0.28]{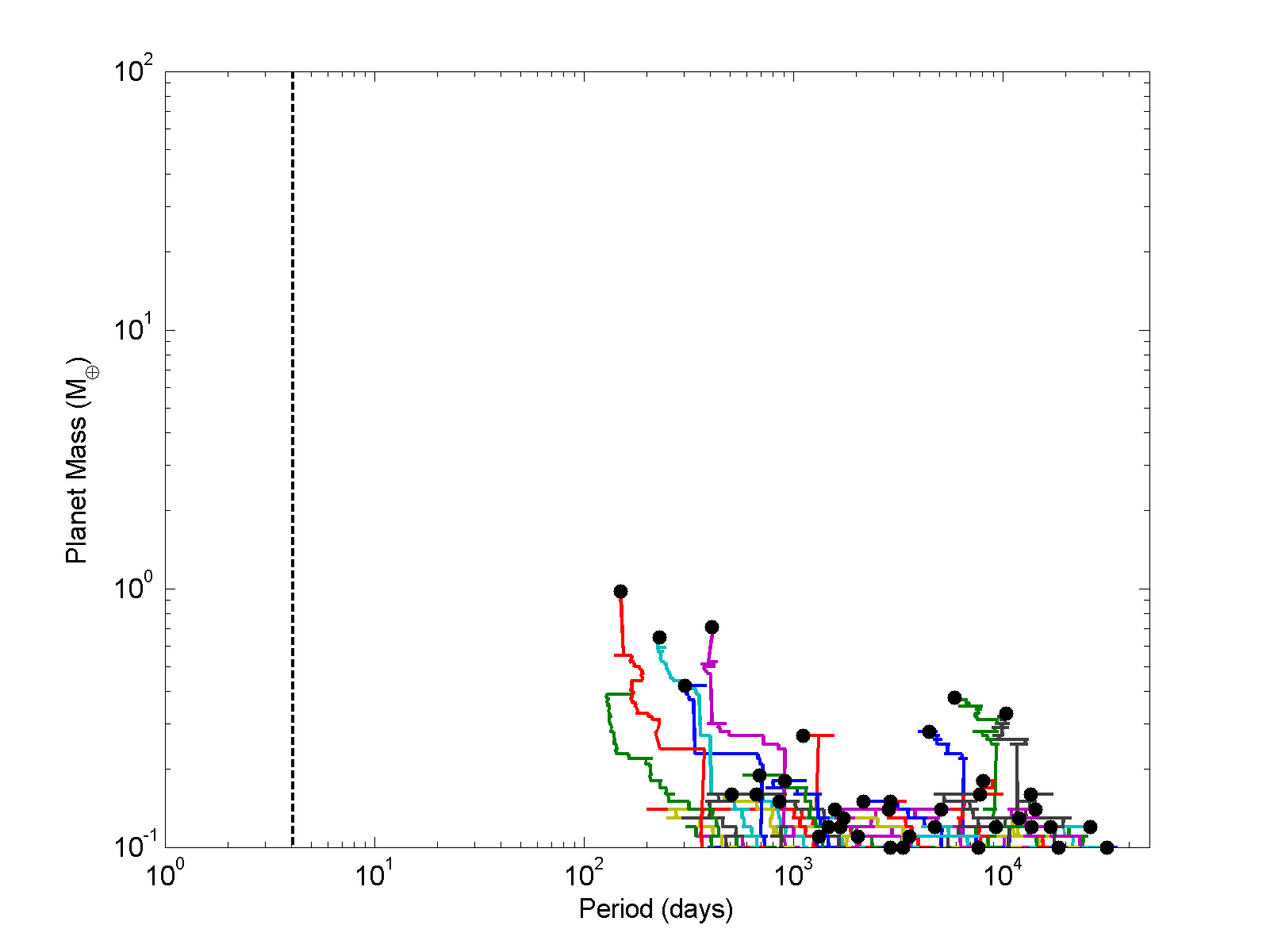}
\caption{Evolution of planet mass versus orbital period
for disc with mass 1 $\times$ MMSN and metallicity 2 $\times$
solar. Left panel: 10~m boulders. Middle panel: 100~m planetesimals. Right panel: 10~km planetesimals.}
\label{fig:MVP-tracks-planetesimal-size}
\end{figure*}

\subsubsection{MGM}
The final states of all runs that exhibited 
\emph{moderate growth and migration} are shown in the middle panel of Figure~\ref{fig:MVP-individual-runs}. Super-Earths and Neptune-mass planets on short period orbits are formed,
and these occur almost always in compact systems (see the lower panels in Figure~\ref{fig:allplanets} in the appendix which shows the final outcomes of all individual runs that were classified as MGM). We note a strong inverse correlation between mass and orbital period in Figure~\ref{fig:MVP-individual-runs} caused by migration.
Low mass planets on short period orbits were shepherded in
as members of resonant convoys driven by more massive planets.
Within individual systems this often led to a direct
correlation between mass and orbital period because migration
was driven by more massive bodies at the outer edge of migrating
resonant chains.

Figure~\ref{fig:MVP-individual-runs} shows that the most
massive survivors have migrated into the magnetospheric cavity.
Their migration was rapid enough to send them in this far,
and they are often accompanied by short-period planets that are surviving members of a resonant convoy that avoided being pushed through the inner boundary.
As mentioned briefly above, runs classified as MGM can be divided into two sub-classes: those that produce objects that migrate quickly enough to reach the magnetospheric cavity, and those which do not, with faster planet growth in more solids-rich discs and/or containing smaller planetesimals/boulders leading to the first sub-class. 

\subsubsection{GFSM}
The final outcomes of runs classified as showing \emph{giant formation and significant migration} are presented in the right panel of Figure~\ref{fig:MVP-individual-runs}. It is clear that all of the surviving gas giant planets have migrated into the magnetospheric cavity, and some of them are accompanied by interior lower mass planetary companions.  

Only models with 10~m boulders and 100~m planetesimals formed giant planets with masses $\ge 35 \me$. All of these planets
except for two were gas-dominated giants - the two exceptions being core-dominated giants (see Table~\ref{tab:plcompo} for definitions). For 10~m boulders
the abundance of solids required to build a gas giant is equivalent to a MMSN disc
with metallicity $1.5 \times$ the solar value. For 100~m 
planetesimals a solids abundance equivalent to a MMSN disc
with metallicity $3 \times$ the solar value is required.
Simulations with 1~km and 10~km planetesimals presented in
CN14 show that giants would have formed in our runs if we had considered disc models with a total solids abundance
equivalent to a MMSN disc with $8 \times$ solar metallicity
(e.g. a $4 \times \mmsn$ disc with $2 \times$ solar metallicity/solids-to-gas ratio.

It is noteworthy
that the most massive surviving (and non surviving) planets
all formed in models with 10~m boulders. Fewer low mass planets are left at large radii in the 100~m planetesimal runs because
planet growth at these radii continues to larger masses in
these runs as the planetesimals do not migrate inwards too
rapidly. This allows the more massive planets formed there to also migrate inwards during the gas disc life time.

\begin{figure*}
\includegraphics[scale=0.28]{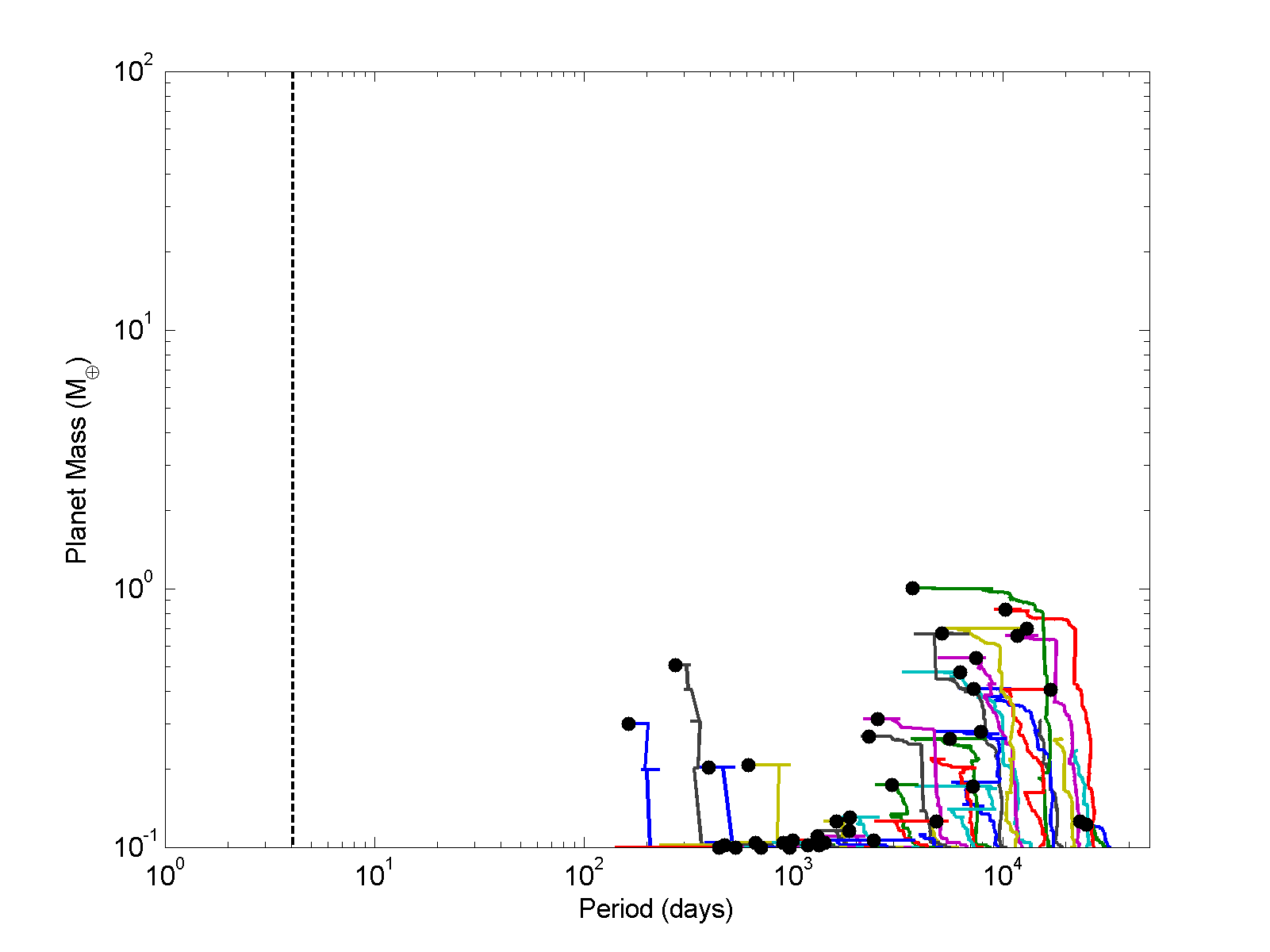}
\includegraphics[scale=0.28]{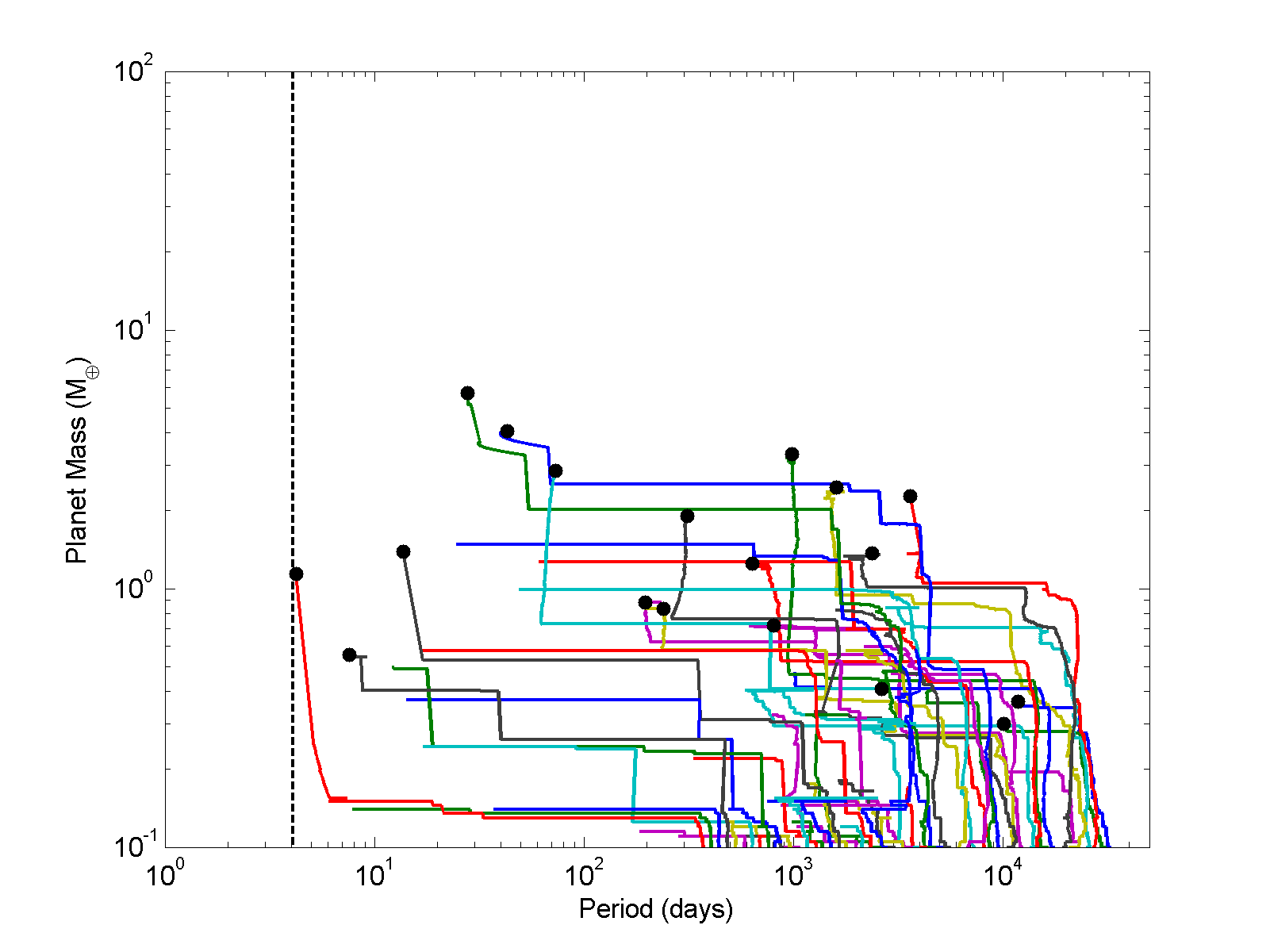}
\includegraphics[scale=0.28]{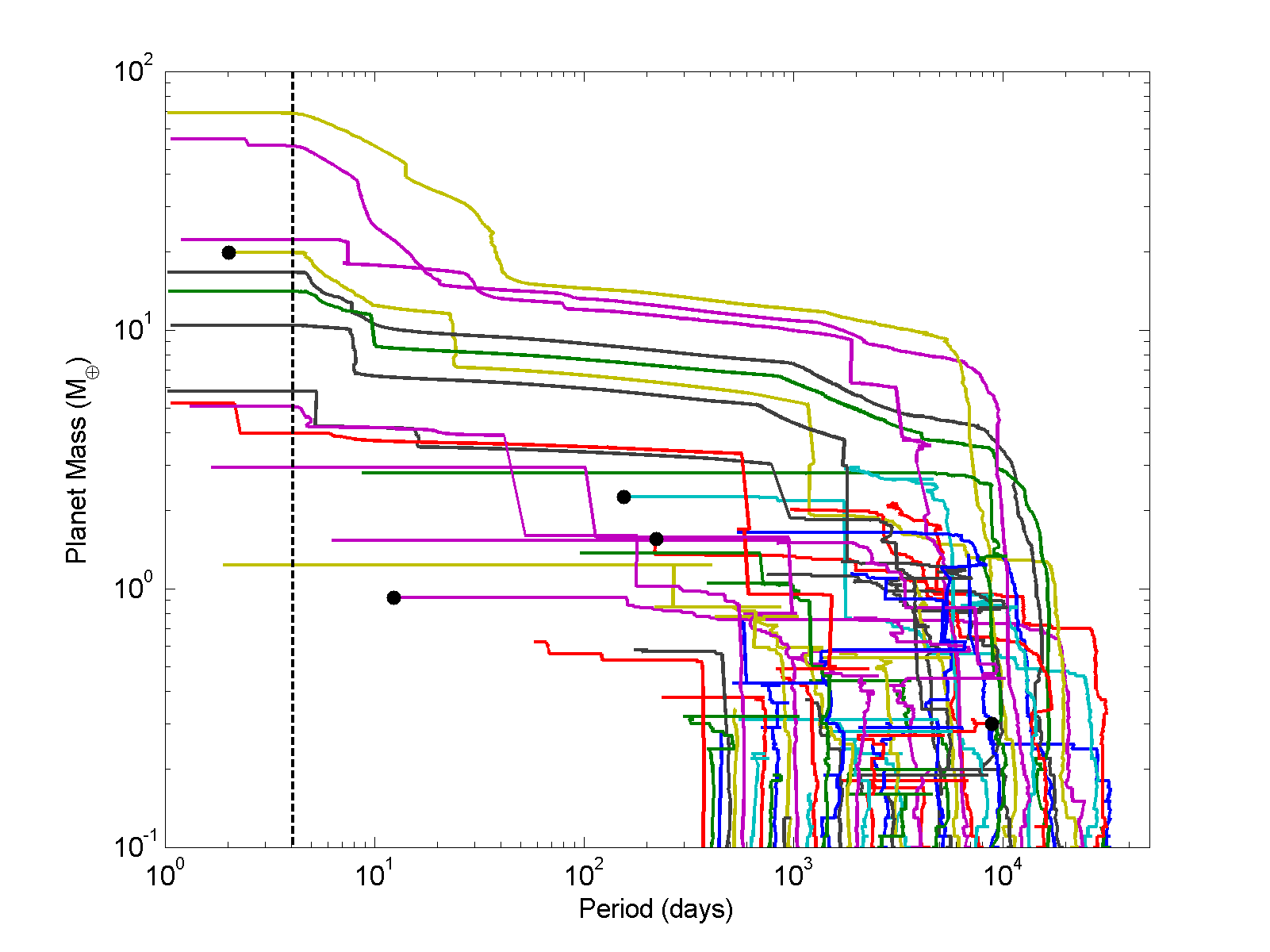}
\caption{Evolution of planet mass versus orbital period
for models with planetesimals sizes of 100~m. 
Left panel: Disc with low solid abundance - run K10.50.1A with
disc mass 1 $\times$ MMSN, metallicity 0.5 $\times$ solar.
Middle panel: Disc with medium-level solid abundance - run
K1.510.1A with disc mass 1.5 $\times$ MMSN, metallicity
1 $\times$ solar. Right panel: Disc with large solid abundance
- run K220.1B with disc mass 2 $\times$ MMSN, metallicity
2 $\times$ solar.}
\label{fig:MVP-tracks-disc mass}
\end{figure*}

\subsection{Evolution as a function of planetesimal radius}
\label{sec:planetesimal-radius}
The simulation results show a very strong dependence on the
planetesimal size adopted, and to highlight this point 
we have plotted planet evolution tracks in the mass--period plane in Figure~\ref{fig:MVP-tracks-planetesimal-size} for simulations with fixed disc properties (disc mass  $1 \times \mmsn$, metallicity $2 \times$ solar) and varying
planetesimal/boulder sizes: 10~m - left panel; 100~m - middle panel; 10~km - right panel. Lines ending in a black filled circle represent the formation of a surviving planet. The left panel shows the formation and rapid inward migration of gas giant planets. The middle panel
shows the formation and inward migration of super-Earths and
Neptune-mass planets. The right panel shows much slower growth
of planets up to approximately one Earth mass and very little 
migration.

\subsection{Evolution as a function of solid abundance}
\label{sec:solid-abundance}
The simulation outcomes show strong dependence on the total mass in solids for a fixed planetesimal size. This is illustrated
in Figure~\ref{fig:MVP-tracks-disc mass}, which shows mass-period
evolution tracks for planets in discs of varying mass and
metallicity for 100~m planetesimals. The left panel shows
results obtained from an anaemic disc with a mass $1 \times \mmsn$ and metallicity $0.5 \times$ solar. Moderate growth and migration is observed in the middle panel for a disc mass of $1.5 \times \mmsn$ and metallicity $1 \times$ solar. The right panel
shows the dramatic change in evolution when the solids abundance
is raised, leading to the formation of numerous Neptune-mass 
and gas giant planets in successive bursts, with a $20 \me$
gas-rich Neptune remaining in the magnetospheric cavity at the
end of the simulation.

\section{Comparison with observations}
\label{sec:comparison}
It is important to re-emphasise that our simulation set does not constitute an attempt at population synthesis. The aim is much simpler: to examine whether or not the model of planet formation and migration presented here is able to form planetary systems similar to those that have been observed within the context of
plausible disc models. We have not used a Monte Carlo approach to select initial conditions from observationally derived distribution functions, and so the frequency with which different types of systems arise in our simulations is not relevant when judging whether or not the planet formation model is successful.
Comparing with observations allows us to determine whether or not the model is capable of producing planets with properties that match those of the observed population (or at least a sub-set of it), and provides a guide for understanding where model improvements are needed.

\subsection{Mass versus period}
\begin{figure}
\includegraphics[scale=0.45]{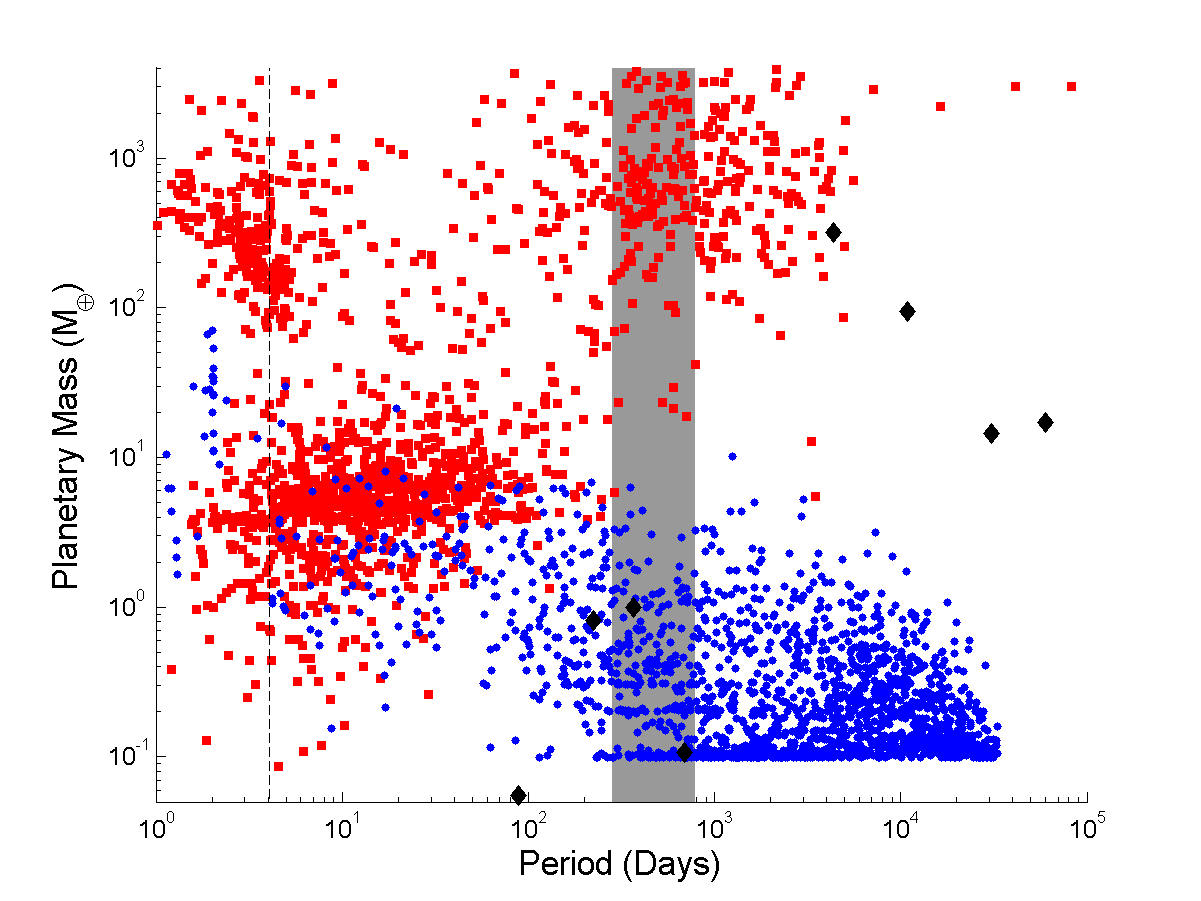}
\caption{Mass vs period plot, comparing observed exoplanets (red squares) 
with our simulation results (blue circles) and 
the Solar System (black diamonds). The dashed line indicates the disc inner edge of $0.05 \au$ 
in our simulations, whilst the grey zone indicates the habitable zone \citep{Kasting}.}
\label{fig:massvperiod}
\end{figure}

Figure \ref{fig:massvperiod} is a mass versus period diagram for the surviving planets from all simulations, along with all confirmed exoplanets \citep{exoplanets_org}.
The vertical dashed line located at $\sim 4$ days shows the position of the disc inner edge in our simulations (i.e the location of the magnetospheric cavity).

The large number of long-period ($>365$ d) low mass planets
($m_{\rm p} \lesssim 5 \me$) produced by the simulations
arises because of the large number of runs that displayed
limited growth (21 out of 36 disc models). These are located in 
a part of the mass-period diagram that is poorly sampled by radial velocity and transit surveys which are biased towards
finding massive planets on short period orbits. Microlensing surveys sample this region of
parameter space and although relatively few planets have been
discovered, constraints obtained from statistical analysis of the data suggest that planets should be common in this region of the diagram \citep{microlensing}.

There is good overlap between the simulation outcomes and
the large numbers of observed short period terrestrial/super-Earth/Neptune-mass planets. In our simulations these planets
tend to form in compact multi-planet systems, similar to
those discovered by \emph{Kepler} \citep{Fabrycky2014} and radial velocity
surveys \citep{2011arXiv1109.2497M}, as discussed in more detail below. The observational data also indicate that there are numerous
systems containing a single planet or which have low multiplicity. The most recent release of \emph{Kepler} data, for example, contains more than 3000 single transiting planet candidates \citep{Mullally2015}. In general, our simulations only produce systems with a short period planet and few objects (if any) orbiting significantly further out when a dominant object (Neptune or gas giant) forms and migrates through the system to the inner cavity. This scenario can clear other planets from the system, leading to low levels of multiplicity. Examples of where this occurred can be seen in Figure~\ref{fig:allplanets}, which shows the final outcomes from all runs with short period planets. 
Forming single planets or low multiplicity systems without a close orbiting dominant body would seem to be difficult in the planet formation scenario presented here, and this may indicate that our choice of inserting 52 planetary embryos at the beginning of the simulations does not match the mode of planet formation occurring most commonly in nature. The prevalence of single or low multiplicity systems may be an indication that planet formation often proceeds by only forming relatively few embryos, in contrast to traditional scenarios of oligarchic and giant impact growth \citep{IdaMakino1993,ChambersWetherill1998}.

The collection of very short period planets ($P < 2$ days) with masses in the range $2 \le m_{\rm p} \le 10 \me$ from the simulations all arose because they migrated into the magnetospheric cavity and were pushed closer to the star by an exterior body that was driving a resonant convoy. These outer planets, that stall finally near the 2:1 resonance with the cavity edge, are also apparent in Figure \ref{fig:massvperiod} and sit in a region of parameter space where there are very few observed planets. We can ascribe these distinct orbital period features in the simulated planet population as being due to adopting a single location for the cavity edge, whereas in reality it will vary from system to system (and with time) due to differences in stellar magnetic field strengths and accretion rates through protoplanetary discs. This will have the effect of blurring the locations of the planets at the 2:1 resonance location and the
interior planets that have been pushed inwards. The group of
more massive planets at 2 days have masses that are not commonly observed, and this may be an indication that our model fails because these bodies should have accreted more gas to become part of the hot-Jupiter population (represented by observed planets with masses $\gtrsim 100 \me$), or should experience substantial evaporation of their atmospheres by stellar X-ray irradiation
on Gyr time scales  \citep{Owen_Jackson_12}, leaving planets with smaller masses in better agreement with observations.
Erosion of the atmosphere through an evaporative wind can also exert a torque on the planet allowing the planet to migrate a few percent of its semimajor axis, if the wind is anisotropic \citep{Teyssandier_owen15}. 

One clear failing in the simulation results is the lack of surviving giant planets with masses $\ge 100 \me$. As mentioned earlier, the most massive planet to form in the simulations
had $m_{\rm p}=160 \me$, but migrated into the star.
The formation of giant planets within our simulation occurred in the inner regions of the disc (orbital radii $\le 1 \au$), and
during times when there were significant amounts of gas remaining.  These giants always migrated into the magnetospheric cavity, before getting trapped at the 2:1 resonance with the disc inner edge. Generally, the last planet that migrated into this region survived, along with a less massive companion if the companion migrated in convoy.
Earlier arriving planets are pushed through the inner boundary of the disc by these late arrivers. The later formation time of these surviving planets causes their masses to be smaller, as the amount of material available for accretion was reduced, explaining why there are not any genuine hot Jupiters or hot Saturns remaining at the ends of the simulations. Once again,
the high multiplicity of our simulated planetary systems may be causing short period giant planets to be removed from the simulations, thus reducing the level of agreement between the models and the observations. In other words, the choice of initial conditions where embryos are equitably distributed throughout the disc may lead to too many planets forming, preventing the survival of early-forming gas giants.

Finally, we note that our models do not even come close to explaining the long period cold-Jupiter population. This is a feature of our simulations that was discussed at length in CN14,
where it was shown that for giant planets to have formed and survived type II migration in our simulations, they would have had to have initiated runaway gas accretion at large orbital radii (typically $> 8 \au$) and during sufficiently late periods of the disc life time when the total disc mass remaining was less than a few tenths of a minimum mass disc.
Forming under these conditions would allow planets to undergo only a moderate amount of type II migration, allowing them to survive at large orbital radii. Trapping giant planet cores at large orbital radii until late times is difficult in our model, however, because the saturation of entropy-related corotation torques leads to rapid inwards type I migration. This point is
illustrated by the migration contours shown in Figure~\ref{fig:contours}.

\begin{figure*}
\includegraphics[scale=0.7]{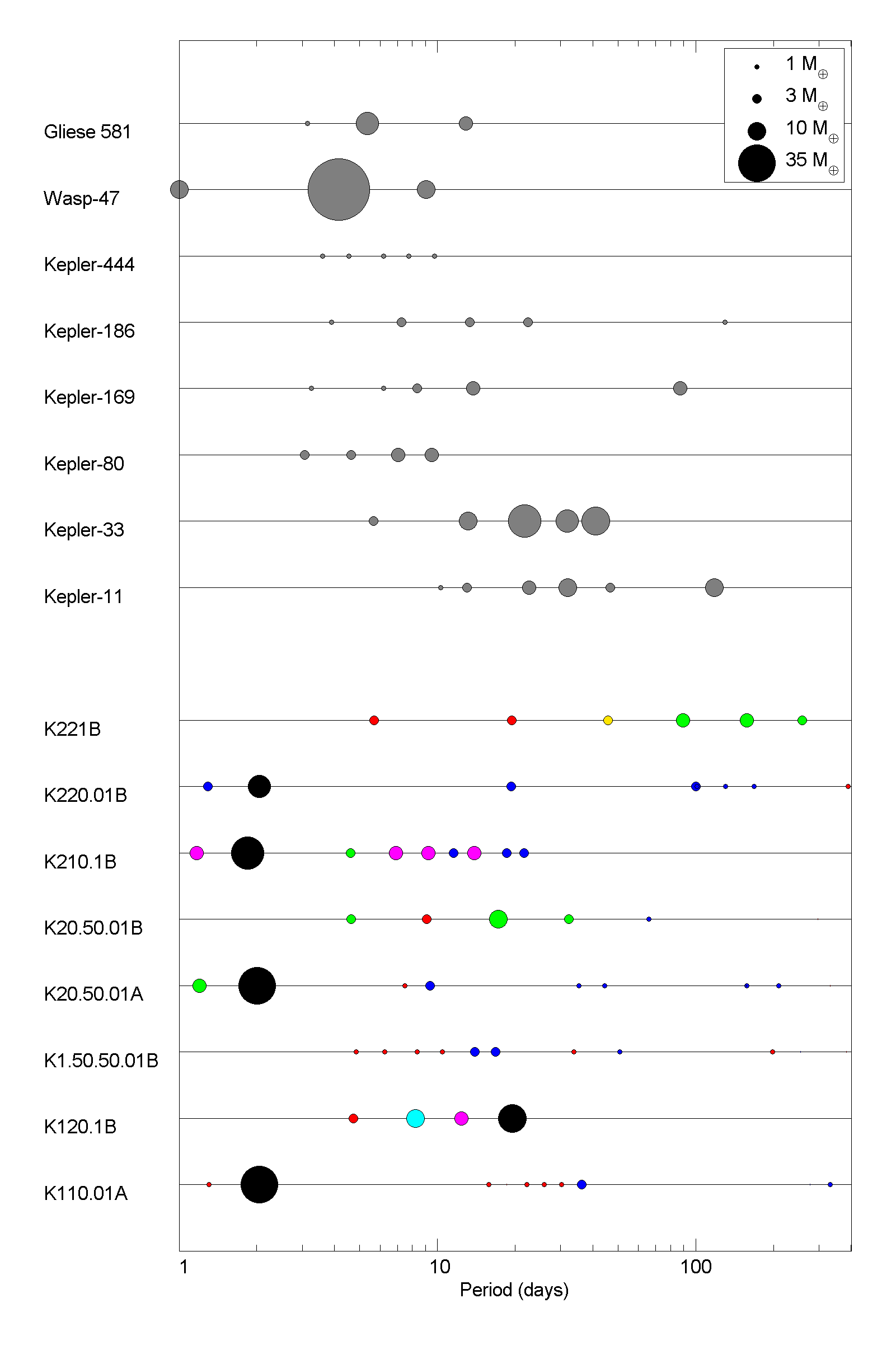}
\caption{Plot comparing observed compact multi-planet systems (upper panels) with simulated systems (lower panels).
Orbital period is indicated on the $x$-axis and planet
masses are indicated by the symbol size (radius scales 
with the square-root of the planet mass) with reference sizes shown in the legend.
Masses for observed systems are either measured masses, or where these are not available they are calculated using the formulae described in \citet{exoplanets_org}. The symbol colours in the lower panels indicate the classification of each planet:
red = rocky terrestrial; blue = water-rich terrestrial; yellow = rocky super-Earth; green = water-rich super-Earth; magenta = mini-Neptune; cyan = gas-poor Neptune; black  = gas-rich Neptune; brown = gas-dominated giant. See Table~\ref{tab:plcompo} for definitions of planet types.}
\label{fig:compsystems}
\end{figure*}
 
 \begin{figure}
\includegraphics[scale=0.45]{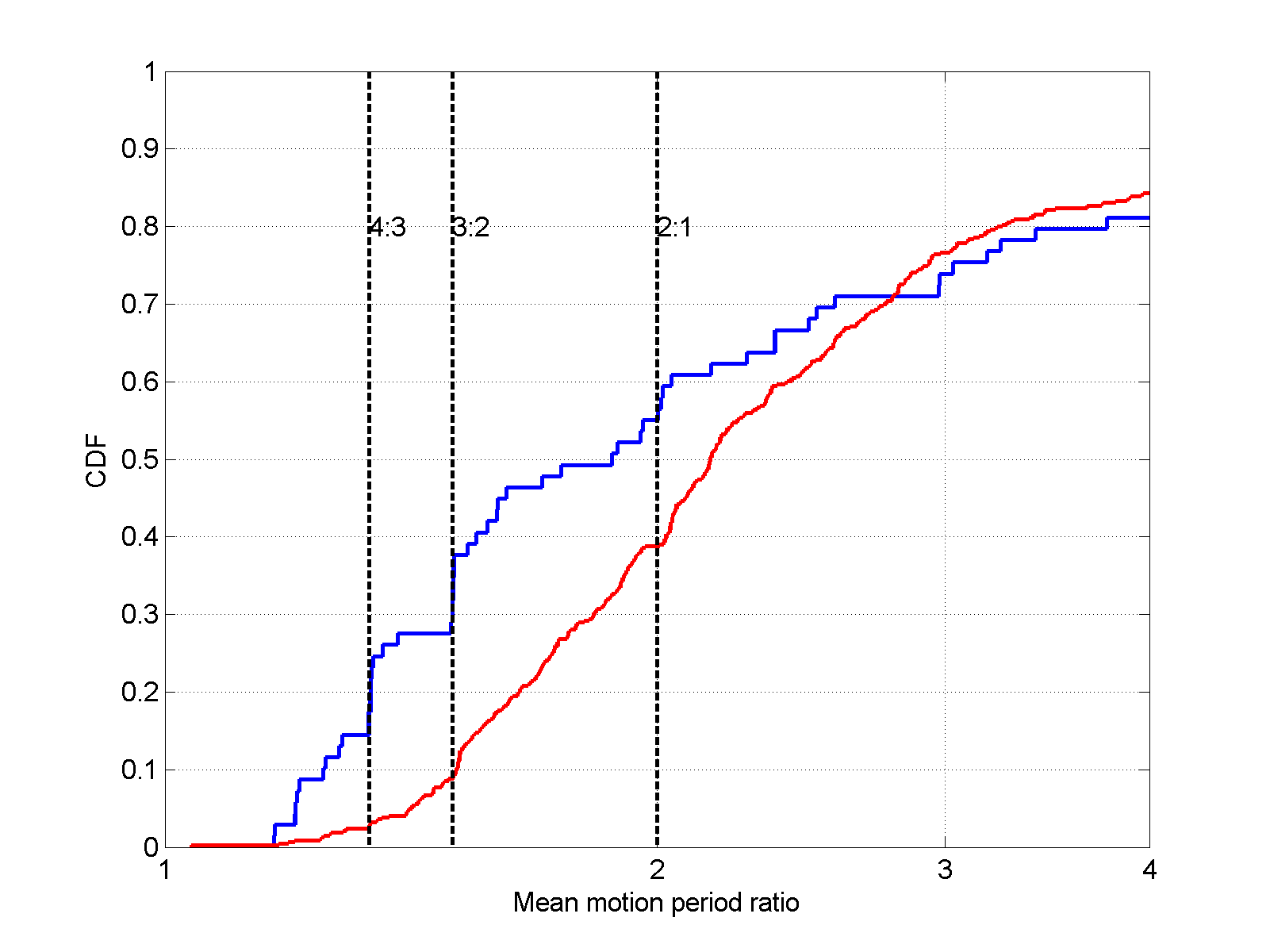}
\caption{Cumulative distribution functions of period ratios between neighbouring planets with periods less than 100 days in the observations (red line) and simulations (blue line).}
\label{fig:resonances}
\end{figure}

\subsection{Comparison with Kepler-like systems}
Figure \ref{fig:compsystems} shows a comparison between a selection of compact Kepler systems, Gliese 581 and Wasp 47 and a selection of our simulated systems. A similar figure is presented in the appendix showing all of the simulated planetary systems that arose from runs resulting in either \emph{moderate growth and migration} or \emph{giant formation and significant migration}. 

Inspection of the simulated planetary systems in Figure~\ref{fig:compsystems} (and Figure~\ref{fig:allplanets} in the appendix) shows that we obtain two basic architectures, one where either a gas-rich Neptune or a gas giant planet has migrated through the system into the inner cavity, and another where the migration has been more modest as planet masses have not grown so massive. The runs K221B, K20.50.01B, K120.1B and K1.50.50.01B displayed the latter type of behaviour, whereas runs K220.01B, K20.50.01A and K110.01A displayed the former type. We obtain outcomes in which the planets are well separated and not in resonance, such as
K221B (for which there was a lot of scattering and growth after the gas disc dispersed) and outcomes such as K1.50.50.01B where the planets are in a chain of resonances at the end of the simulation. Note that Figure~\ref{fig:allplanets} shows which pairs of planets in the final systems are in mean motion resonances. We also find a small number of coorbital planets at the end of the runs (3 trojan systems and 1 horseshoe system were found orbiting within 200 days across all runs. These systems are shown as being in 1:1 resonance in Figure~\ref{fig:allplanets}). All coorbital planets were found in systems where at least one planet underwent rapid and large scale migration, causing bodies to be scattered onto eccentric orbits that quickly damped once the rapid migrator had passed through the system. This concurs with previous studies of coorbital planet formation which showed that these bodies are a direct consequence of violent relaxation in a highly dissipative environment \citep{CresswellNelson2006}.

While it is difficult to perform a quantitative comparison between the simulated and the observed planets, certain similarities can be noted. For example, Kepler 444 looks similar to the inner four planets of K1.50.50.01B. These four rocky-terrestrial planets were shepherded in by the exterior more massive water-rich terrestrials, and hence formed a resonant convoy. This is one way in which the Kepler 444 planets could have arrived at their observed locations and provides an alternative to \emph{in situ} formation (but relies on there being a more massive, undetected planet orbiting further from the star). Kepler 169, 186 and 80 look similar to K20.50.01B, and Kepler 11 and 33 have broad similarities with K120.1B. Although the Kepler sample does not contain examples of compact multi-systems with massive, short period planets (perhaps because these are more dynamically disturbed and therefore not transiting or close to resonances such that they are detectable through transit timing variations), Gliese 581 and Wasp 47 provide two examples that have architectures similar to K210.1B and K220.01B.

\subsection{Period ratios and orbital spacings}
\label{sec:periodratios}
Figure~\ref{fig:resonances} compares the cumulative distributions of period ratios between neighbouring planets with masses $\ge 1 \me$ and orbital periods less than 100 days obtained from the simulations (upper blue curve) and the Kepler systems (lower red curve). The sample of Kepler planets was defined by choosing bodies with orbital periods $\le 100$ days and radii $\ge 1 {\rm R}_{\oplus}$. This lower radius limit was adopted to account for possible incompleteness in the Kepler sample for planets with small radii. It is clear that the simulated systems are generally more closely packed after run times of 10 Myr, and the structure observed in the distribution shows that this is due in part to there being a number of planet pairs in resonance. The step-like features in the plot show that the 7:6, 6:5, 5:4, 4:3, 3:2 and 2:1 resonances are occupied. Whereas just an isolated pair of migrating planets are
likely to be trapped in either the 2:1 or 3:2 resonances if they undergo smooth migration \citep{Paardekooper2013}, we find that migration in a crowded system allows diffusion through successive resonances to
occur such that high degree resonances can be occupied,
in agreement with earlier studies by \cite{CresswellNelson2006, cressnels}. Although resonant systems are relatively rare in the Kepler data, it is worth noting that Kepler 36 has two planets very close to the 7:6 resonance \citep{Carter2012,Paardekooper2013},
and some of the planet pairs in Kepler 444 are reported to be in 5:4 \citep{Campante2015}. Other examples of systems in resonance or near resonance, including 3 body resonances and resonant chains, are Kepler 50 (6:5), Kepler 60 (5:4, 4:3) \citep{Steffen2012}, Kepler 221 (3 body resonance where the mean motion combination $2 n_{\rm in} - 5 n_{\rm mid} + 3 n_{\rm out}$ has been found to librate around 180 degrees) \citep{Fabrycky2014}.
\begin{figure}
\includegraphics[scale=0.45]{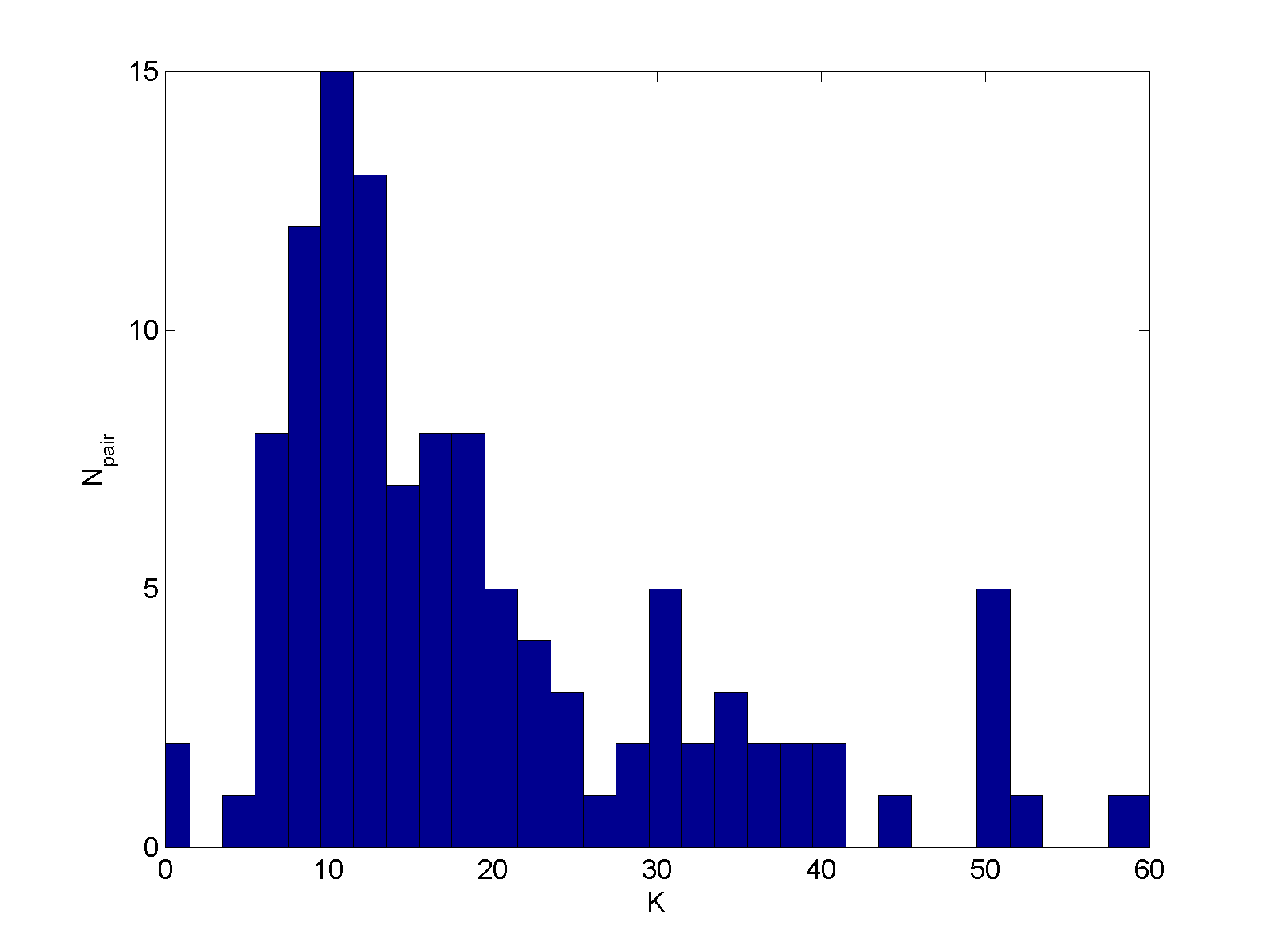}
\caption{Histogram showing the distribution of separations between neighbouring planets with masses $\ge 1 \me$, measured in units of the mutual Hill radius.}
\label{fig:separations}
\end{figure}
Furthermore, it has been noted in numerous studies \citep[e.g.][]{Fabrycky2014} that the distribution of planet period ratios contains an excess of planets just outside of 3:2 and 2:1, suggesting that the resonances have been dynamically important during the evolution but may have been broken by stochastic migration in a turbulent disc \citep{Adams2008,ReinPapaloizou2009}, by tidal interaction with the central star \citep{TerquemPapaloizou2007}, by orbital repulsion due to damping of nonlinear spiral waves \citep{BaruteauPapaloizou2013}, 
by overstability in librations about resonant centres \citep{GoldreichSchlichting2014}\textbf{, or because of scattering due to interactions with or accretion of residual planetesimals \citep{ChatterjeeFord15}. We observe that in a handful of simulations, planetesimal scattering after full gas disc dispersal does occur, breaking mean-motion resonances between neighbouring planets, in agreement with \citet{ChatterjeeFord15}.}
It is noteworthy that a number of the compact systems are orbiting in regions where their nascent protoplanetary discs are expected to have sustained MRI turbulence due to the local temperature being in excess of 1000~K \citep{UmebayashiNakano1988}, and so may have been subjected to stochastic forcing of their orbits while the gas disc was present. To seek evidence for this transition to turbulence we have examined the minimum periods of planets in the compact Kepler multi-systems to see if they correlate with the effective temperature of the host star, but there is
no evidence of a correlation. At present there is no clear evidence that the transition to turbulence in the inner regions of the protoplanetary discs that formed the Kepler systems played a decisive role in dynamically shaping these systems. 
     
It is possible that a number of our simulated systems may be dynamically unstable on time scales much longer than the 10 Myr run times, such that subsequent mutual collisions increase separations between adjacent planets. In a recent study, \cite{PuWu2015} used N-body simulations to show that compact Kepler-like multi-planet systems tend to remain stable for Gyr time scales only if the typical mutual separation between neighbouring planets is approximately 12 mutual Hill radii. Figure~\ref{fig:separations} shows the distribution of separations between neighbouring planets present at the end of the simulations, and while many planet pairs are well separated there are a significant number whose orbital spacings may be too small for long term stability. Running the simulations for long enough to test this goes beyond the scope of this paper, but will be studied in future work as it may be the case that the mean motion resonances discussed above provide protection against instability. We note that the objects with period ratios of unity shown in Figure~\ref{fig:separations} are the coorbital planets mentioned previously.

\section{Discussion and conclusion}
\label{sec:discussion}
We have implemented a model of planet formation based on a scenario in which numerous planetary embryos are distributed across a wide range of semimajor axes, embedded in a sea of boulders or planetesimals that act as the primary feedstock for planetary growth. The model has a comprehensive list of ingredients: planetary embryo growth through boulder/planetesimal accretion and mutual collisions; a 1D viscous gas disc model, subject to irradiation from the central star and a photoevaporative wind; type I migration using the most up-to-date prescriptions for Lindblad and corotation torques; a transition to gap formation and type II migration when gap formation criteria are satisfied; gas accretion onto solid cores. The disc has an increase in viscosity where the temperature $T > 1000$~K, to mimic unquenched MHD turbulence developing in the inner disc, and a magnetospheric cavity that creates an inner edge in the gas disc at an orbital period of 4 days. The aim of this study is to determine which types of planetary systems emerge from the planet formation model as a function of disc parameters (mass and metallicity) and planetesimal/boulder sizes. 
The main results from our simulations can be summarised as follows.

(i) System evolution can be categorised into three distinct modes that depend on the total amount of solids present in the disc and the sizes of the boulders/planetesimals. \\
- When planetesimal/boulder radii are small ($\le 100$~m) \emph{limited planetary growth} arises when the inventory of solids is small. When planetesimal radii are large ($\ge 1$~km), limited growth arises for all discs models considered, except the one that is the most massive and solids-rich. Planets with maximum masses $\sim 3 \me$ form during the gas disc life times, and show only very modest migration.\\
- \emph{Moderate growth and migration} arises in only the most solids-rich disc considered when planetesimal sizes are 1~km, and for disc models with intermediate abundances of solids when the planetesimal/boulder sizes $\le 100$~m. Planets are able to grow to super-Earth or Neptune masses during the disc life time, and may undergo large-scale migration.\\
- \emph{Giant formation and significant migration} is observed in the most solids-abundant discs when boulder/planetesimal sizes were $\le 100$~m, but did not arise in any of the runs with larger planetesimals. Generally, multiple episodes of planet formation occur, and gas giant planets with masses $\ge 35 \me$ form and undergo large scale migration before stalling in the magnetospheric cavity.
The final surviving short period planets are normally the
last ones to arrive in the magnetospheric cavity, with the earlier arrivals being pushed through the inner boundary by the planets that arrive there later.

(ii) Considering systems of short-period planets, we can identify two basic architectures that emerge from the simulations. The first normally consists of a combination of terrestrial planets, super-Earths and low mass Neptunes, where no planet managed to migrate into the magnetospheric cavity. The shortest period orbits in these systems are normally 4 -5 days. The second architecture consists of at least one dominant planet (a gas giant or a relatively massive Neptune) that migrated and stalled in the magnetospheric cavity with a period of $\sim 2$ days. In approximately 50\% of cases, this planet has an interior companion (terrestrial planet, super-Earth or Neptune) which is almost never in resonance because of dynamical interactions and collisions with other planets during the evolution. In most cases where a dominant short period planet formed, there are a number of exterior planets orbiting with periods in the range $5 \lesssim P \lesssim 80$ days.

(iii) The planetary systems display a range of heterogeneity in composition versus orbital period. Systems that formed under relatively quiescent conditions, without a rapidly migrating gas giant or Neptune, have rocky
bodies orbiting interior to volatile rich bodies. Systems 
that contained rapidly migrating giants or Neptunes, that end up in 2 day orbits, often experienced significant scattering, and these systems can have rocky bodies in exterior orbits in close proximity to volatile-rich bodies.

(iv) The planetary systems that emerge from the simulations tend to be closer packed than the observed Kepler systems. The most common spacing between neighbouring planets is
10 - 12 mutual Hill radii, and \citet{PuWu2015} have shown that such systems are likely stable over Gyr time scales.
There are, however, numerous simulated planet pairs where the ratio of spacing to mutual Hill radius $< 10$, and these might cause the systems to evolve and change their spacing through collisions if evolved beyond the 10~Myr that we have considered, improving the agreement with observations. We note, however, that mean motion resonances may help stabilise our simulated systems compared with those considered by \citet{PuWu2015}.

(v) One reason for the difference in the distributions of observed versus simulated period ratios is that mean motion resonances are common among our final planetary systems.
We find examples of 7:6, 6:5, 5:4, 4:3, 3:2 and 2:1, with the latter three resonances being rather common. It is well known that most of the compact Kepler systems do not display mean motion resonances, even though there is evidence for the 2:1 and 3:2 resonances having been dynamically important in the past, and a few individual systems appear to host resonant pairs or triples. One possible explanation for the greater numbers of resonant systems arising in the simulations is the neglect of stochastic forces in the inner disc regions due to MHD turbulence \citep{NelsonPapaloizou2004,Nelson2005} which can cause planets to diffuse out of resonance \citep{Adams2008,ReinPapaloizou2009}. It remains to be seen whether or not inclusion of this effect can increase the agreement between observations and theory on the frequency of mean motion resonances. One further point worthy of note is that the frequency of resonances arising in the simulations is higher for those architectures that contain a dominant planet orbiting with a 2 day period. Systems without a dominant short period planet underwent more quiescent evolution during the gas disc life time, but also experienced more scattering after removal of the disc and this leads to systems that contain few resonances (see Figure~\ref{fig:allplanets}). Thus, it is important to note that there is a mode of planet formation that includes large scale migration but which does not result in systems that are members of resonant chains.

(vi) A number of coorbital planets were formed in our simulations (three trojan systems, and one undergoing mutual horseshoe orbits, were found to orbit with periods $< 200$ days). These all formed in systems where at least one dominant planet underwent migration through the planetary swarm, causing large amounts of scattering. In earlier work \citet{CresswellNelson2006, cressnels} have shown that coorbital planets arise as a consequence of violent relaxation in crowded planetary systems with strong eccentricity damping,
and our results are in agreement with these earlier findings.

(vii) Numerous gas giant planets were formed in our simulations, and some survived after migrating into the magnetospheric cavity. The most massive planet to form 
was a $160 \me$ gas giant, but this was pushed through the inner boundary of the computational domain by a planet that arrived in the magnetospheric cavity at a later time. The most massive surviving planet was a $70 \me$  ``hot Saturn" on a 2 day orbit. CN14 undertook a detailed examination of the conditions required for the formation and survival of longer period giant planets against type II migration, and showed that a Jovian mass planet halting its migration at $5\au$ needs to start runaway gas accretion and type II migration at a distance of $\sim 15 \au$ from the central star. This has not occurred in any of our simulations (this paper, or CN14, or in the many low resolution test simulations that we have run and not published), because of the difficulty of forming a core and keeping it at such large orbital radius. We have concluded that forming and retaining long period giant planets requires a set of disc conditions that are quite different from those that we have considered thus far. A potential solution to the problem will be presented in a forthcoming paper (Coleman \& Nelson in prep.).

\subsection{Formation of Kepler 444 and 42}
The Kepler 444 and 42 systems are examples of short period compact low mass planetary systems. All have radii substantially smaller than the Earth's. Kepler 444 is a 5-planet system orbiting a $0.76$~M$_{\odot}$ K0V star with ${\rm [Fe/H]} \sim -0.55$, where the innermost orbital period is 3.6 days and the outer planets are close to the 5:4, 4:3, 5:4 and 5:4 mean motion resonances \citep{Campante2015}. Kepler 42 is a 3-planet system orbiting a $0.13$~M$_{\odot}$ M3V star with ${\rm [Fe/H]} \sim -0.3$. Orbital periods are 0.453, 1.214 and 1.865 days \citep{Muirhead2012}, so there are no first-order mean motion resonances. We showed in Sect~\ref{sec:results} that planet masses need to be in excess of $\sim 3 \me$ for migration over large distances to be effective, and given the low metallicities of these systems they are most likely explained by \emph{in situ} formation after delivery of solids through drag-induced drift into the disc inner regions. Although large scale migration of these
planets is implausible, the resonant or near-resonant configuration of the Kepler 444 planets suggests that modest migration may have occurred. The outermost planet being the largest (and presumably most massive) would lead to the necessary convergent migration.  

\subsection{Formation of short period super-Earths in low metallicity discs}
Our simulations demonstrate how difficult it is to grow planets that are massive enough to undergo significant type I migration during the gas disc life time when growth is dominated by the accretion of large ($\ge 1$~km) planetesimals in discs with a moderate inventory of solids. This is because growth time scales are slow for large planetesimals. In addition, if a planet approaches its local isolation mass it will not be massive enough to migrate such that it can accrete from undepleted sources of planetesimals. The situation becomes more difficult in a low metallicity environment, and the existence of short period super-Earths around stars such as Kapteyn's star \citep{Anglada-Escude2014}, Gliese 581  \citep{Udry2007}, HD 175607 \citep{Mortier2015} and the numerous low-metallicity hosts of Kepler systems \citep{Buchhave2014} suggests that these planets did not form via the classical oligarchic growth picture of widely distributed embryos accreting from a swarm of large planetesimals. These systems instead point towards planetary embryos growing into type I migrating super-Earths by accreting from a supply of highly mobile small planetesimals, boulders or pebbles \citep[e.g.][]{OrmelKlahr2010, Lambrechts12}, as this is the only means available of exceeding local isolation masses. On the other hand, the requirement for the local solids-to-gas ratio to be approximately twice solar in order for the streaming instability to operate and generate large planetesimals that can acts as the seeds of growing planets \citep{Johansen2009} suggests that small particles must first concentrate in specific disc regions due to the existence of zonal flows \citep{JohansenYoudin2009,Bai2014}, vortices \citep{FromangNelson2005} or dead zone interfaces \citep{Lyra2009} in order to create local enhancements of solids. Such a collect-and-grow scenario would appear to offer the best hope for explaining the existence of planets in the lowest metallicity environments.    

\subsection{Future work and directions}
The long term aim of this project is two-fold: to construct a simulation tool for modelling planet formation that comprises accurate prescriptions for the important ingredients for planet building and evolution; to determine whether or not it is possible to explain the diversity of known planetary systems using a comprehensive model of planet formation, loosely based on the classical core accretion model, operating under different initial conditions and environments. A particular issue of interest is explaining the known population of gas giant planets, and we will present a study of this in a forthcoming paper. Areas of future improvement to our model include:\\
(i) Calculation of gas envelope accretion using self-consistent computations that take account of the changing local nebula conditions, rather than using fits to the \citet{Movs} models as is done now. The atmosphere models of \citet{PapNelson2005} are being incorporated into the code, and results from these calculations will be presented in a forthcoming publication. \\
(ii) Improving the disc model so that stellar irradiation of the disc inner regions is treated more accurately. Our 1D treatment of stellar irradiation underestimates the level of heating near the star, and this allows the temperature to fall below 1000~K everywhere in the disc at late times, such that no region of the disc maintains fully developed MRI turbulence. A more realistic treatment would allow the temperature to always be above 1000~K out to a radius $R_{1000} \sim R_* (T_*/{\rm 1000 \,K})^2$, which for our model corresponds to a distance of $0.17 \au$ from the star.\\
(iii) Include the effects of stochastic migration when planets and planetesimals enter disc regions where $T \ge 1000$~K. This will influence the ability of planet pairs to maintain mean motion resonances. \\
(iv) Improve the migration model by including 3D effects \citep{Fung2015} and the influence of the planet luminosity \citet{Benitez-LlambayMasset2015}. \\
(v) Small boulders and planetesimals are able to migrate inwards from beyond the snow line, and in principle these should sublimate quite rapidly. We have not included submlimation in our models, and analysis of the results indicates that planets accreting icy planetesimals that have migrated interior of the ice line increase their masses by at most $3 \%$. Nonetheless a model of planetesimal sublimation should be included for self-consistency.

Simulations incorporating these improvements will be presented in future publications in order to determine how these modifications change the simulation outcomes. Once a suitably sophisticated model of planet formation has been constructed it will be used to generate a synthetic planet population based on initial conditions and physical parameters drawn from observational constraints to determine the level of agreement with exoplanet observations.

\appendix
\section{Presentation of all simulated compact systems}
\begin{figure*}
\includegraphics[scale=0.7]{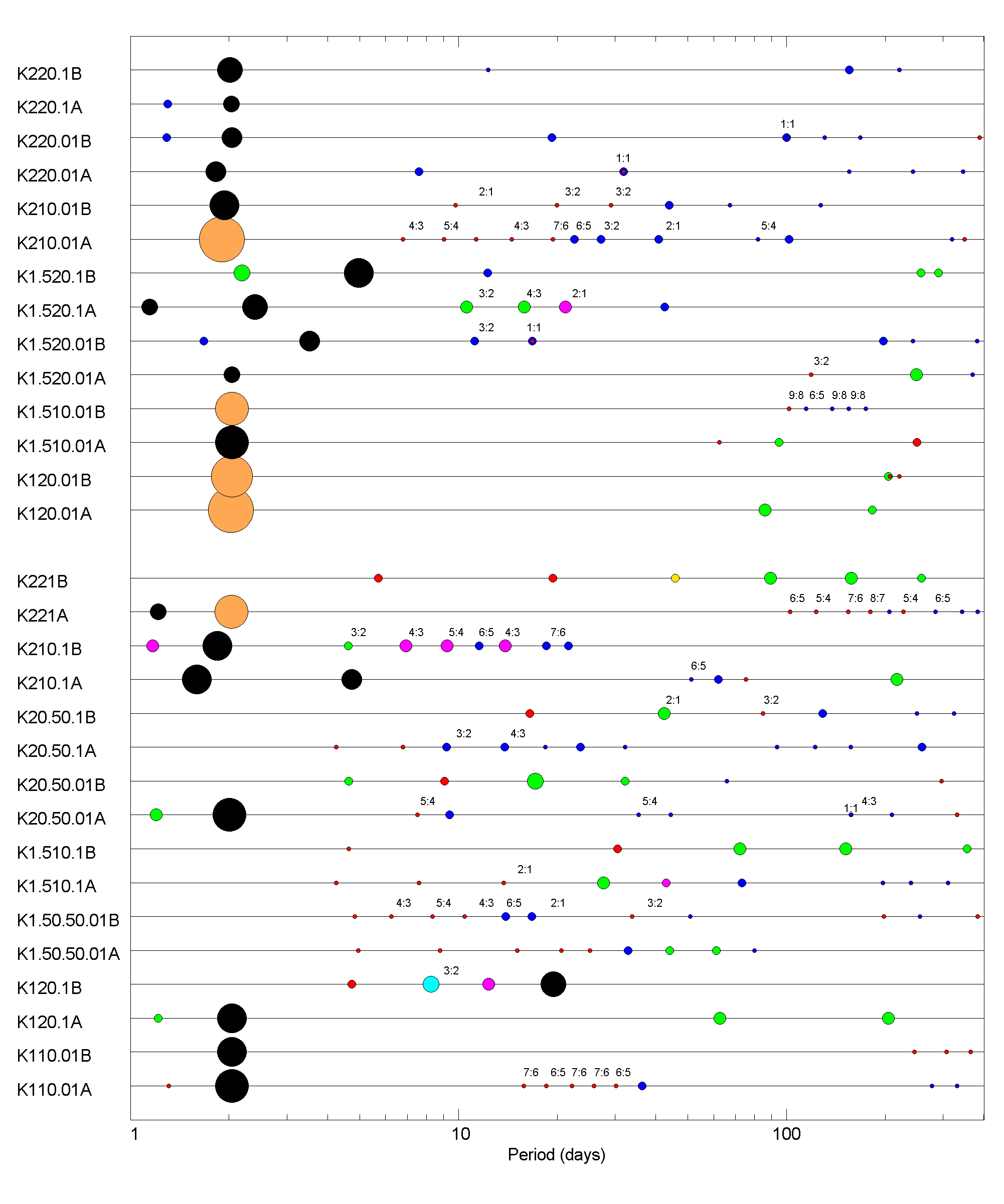}
\caption{Plot showing all compact multi-planet systems formed in the simulations. The upper panels represent planets formed in runs classified as \emph{giant formation and significant migration}, the low panels show outcomes from \emph{moderate growth and migration} runs.
Orbital periods are indicated on the $x$-axis and planet
masses are indicated by the symbol size, as in Figure \ref{fig:compsystems}.
The symbol colours indicate the classification of each planet: red = rocky terrestrial; blue = water-rich terrestrial; yellow = rocky super-Earth; green = water-rich super-Earth; magenta = mini-Neptune; cyan = gas-poor Neptune; black  = gas-rich Neptune; brown = gas-dominated giant. See Table~\ref{tab:plcompo} for definitions of planet types.}
\label{fig:allplanets}
\end{figure*}

Figure~\ref{fig:allplanets} shows all of the compact systems that formed and survived in the simulations. The planets shown in the upper panels all formed in simulations classified as \emph{giant formation and significant migration}, and the rest formed in simulations classified as \emph{moderate growth and migration}. Pairs of planets that are coorbital or are in first order mean motion resonances are indicated by the integers printed above and between the relevant pair.

\bibliographystyle{mnras}
\bibliography{references}{}
\label{lastpage}
\end{document}